\documentclass[12pt]{article}
\usepackage {amsmath}
\usepackage[dvips]{graphicx}
\usepackage{amssymb}
\usepackage{cite}
\newcommand{\be}{\begin{equation}}
\newcommand{\ee}{\end{equation}}
\newcommand{\bear}{\begin{eqnarray}}
\newcommand{\eear}{\end{eqnarray}}
\newcommand{\tanb}{\tan \beta}
\newcommand{\tb}{\textbf}
\newcommand{\gl}[1]{\lambda_#1}
\newcommand{\gx}[2]{\chi _{#1} \left( {#2 } \right)}
\newcommand{\fe}[2]{f_{#1} \left( {#2 } \right)}
 
\newcommand{\bra}[1]{\left\langle #1\right|} 
\newcommand{\ket}[1]{\left| #1\right\rangle} 
\newcommand{\vev}[1]{\left\langle #1\right\rangle}

\newcommand{\goes}{\rightarrow} 
 
\newcommand{\GeV}{\; \mathrm{GeV}} 
\newcommand{\TeV}{\; \mathrm{TeV}} 
\newcommand{\lapproxeq}{\lower .7ex\hbox{$\;\stackrel{\textstyle  
<}{\sim}\;$}} 
\newcommand{\gapproxeq}{\lower .7ex\hbox{$\;\stackrel{\textstyle  
>}{\sim}\;$}} 
\newcommand{\stackdown}[2]{\lower 1.4ex\hbox{$\;\stackrel{\textstyle{#1}}  
{\scriptstyle{#2}}\;$}}
\newcommand{\beq}{\begin{equation}} 
\newcommand{\eeq}{\end{equation}} 
\newcommand{\lsp}{\tilde{\chi}}
\newcommand{\mlsp}{m_{\lsp}}
\newcommand{\xsec}{\vev{ \sigma v_{rel} }}

\newcommand{\relic}{\Omega_{\lsp}\,h_0^2}

\newcommand{\too}{\longrightarrow}
\newcommand{\bea}{\begin{eqnarray}}
\newcommand{\eea}{\end{eqnarray}}

\makeatletter 
\def\slash{\@ifnextchar[{\fmsl@sh}{\fmsl@sh[0mu]}} 
\def\fmsl@sh[#1]#2{%
  \mathchoice 
    {\@fmsl@sh\displaystyle{#1}{#2}}%
    {\@fmsl@sh\textstyle{#1}{#2}}%
    {\@fmsl@sh\scriptstyle{#1}{#2}}%
    {\@fmsl@sh\scriptscriptstyle{#1}{#2}}} 
\def\@fmsl@sh#1#2#3{\m@th\ooalign{$\hfil#1\mkern#2/\hfil$\crcr$#1#3$}} 
\makeatother 

\begin{document} 
\begin{titlepage} 
 
\begin{flushright} 
\parbox{7.4cm}{hep-ph/0404286\\
               MIFP-04-08, ACT-03-04\\
               UA-NPPS/BSM-04/02\\
               UMN-TH-2305/04, FTPI-MINN-04/16}
\end{flushright} 
\vspace*{5mm} 
\begin{center} 
{\large{\textbf {Partial wave treatment of Supersymmetric Dark Matter 
in the presence of $CP$-violation 
}}}\\
\vspace{11mm} 
{\bf M.~\ Argyrou} $^{1}$, {\bf A.~B.\ Lahanas} $^{1}$,
{\bf D.~V.~Nanopoulos} $^{2}$ and 
{\bf V.~C.\ Spanos} $^{3}$   

\vspace*{6mm} 
 $^{1}$ {\it University of Athens, Physics Department,  
Nuclear and Particle Physics Section,\\  
GR--15771  Athens, Greece}

\vspace*{6mm} 
$^{2}$ {\it George P. and Cynthia W. Mitchell 
                Institute of Fundamental Physics, \\   
     Texas A \& M University, College Station,  
     TX~77843-4242, USA \\[1mm]
     Astroparticle Physics Group, Houston 
     Advanced Research Center (HARC),\\ 
     The Mitchell Campus, The Woodlands, TX 77381, USA  \\[1mm] 
     Chair of Theoretical Physics,  
     Academy of Athens,  
     Division of Natural Sciences, 28~Panepistimiou Avenue,  
     Athens 10679, Greece }

\vspace*{6mm} 
 $^{3}${\it William I. Fine Theoretical Physics Institute,
University of Minnesota,  \\ Minneapolis, MN 55455, USA}

\end{center} 
\vspace*{3mm} 
\begin{abstract}
We present an improved partial wave analysis of the dominant
LSP annihilation channel to a fermion-antifermion pair which avoids the 
non--relativistic expansion being therefore applicable near thresholds and 
poles. The method we develop allows of contributions of any partial wave in
 the total angular momentum $J$ in contrast to partial wave analyses in terms 
of the orbital angular momentum $L$ of the initial state, which is usually 
truncated to p-waves, and yields very accurate results. 
The method is formulated in such a way as to allow easy handling 
of $CP$-violating phases residing in supersymmetric parameters.  
We apply this refined partial wave technique in order 
to calculate the neutralino relic
density in the constrained MSSM (CMSSM) in the presence of $CP$-violating 
terms occurring in the Higgs - mixing parameter $\mu$ and 
trilinear $A$ coupling for large
$\tan \beta$. The inclusion of $CP$-violating phases in $\mu$ and 
$A$  does not upset significantly
the picture and the annihilation of the LSP's
to a $b \bar{b}$,  through Higgs exchange, is still
the dominant mechanism in obtaining cosmologically
acceptable neutralino relic densities in 
regions far from the stau-coannihilation
and the `focus point'. Significant changes can occur if we allow 
for phases in the gaugino masses 
and in particular the gluino mass.  

\end{abstract} 
\end{titlepage} 
%
\baselineskip=18pt 
\section{Introduction}

In the framework of Supersymmetric Theories, in which R - parity is
conserved, the lightest of the neutralinos, $\lsp$, which
is also the lightest supersymmetric particle (LSP) is 
a promising candidate for cold  Dark Matter (DM) \cite{EHNOS}.

The knowledge of its relic density today combined with recent experimental 
observations, especially 
those coming from WMAP \cite{wmap,LNM}, put severe
constraints on the parameters
of the theory which if combined with other data enhance the possibilities
of discovering Supersymmetry
in future accelerator experiments.

In the framework of the Constrained Minimal 
Supersymmetric Standard Model 
small values of the neutralino relic density, $\relic$, in the vicinity of
$ 10 \% $ range as recent astrophysical data on Cold Dark Matter
suggest~\cite{eoss,LN,cmssmmap},\cite{LNM}, are obtained in regions 
of the parameter space which belong to
either the neutralino - stau coannihilation region \cite{coanni1,coanni2}, 
or to the focus point~\cite{hbrsb,focus},  
or to a region in the neighborhood of the line along which 
the pseudoscalar Higgs has a mass approximately 
twice that of the LSP mass \cite{Drees,LNS1,gannis,baer,LNS4,LS,LN},  
which is characterized by large $ \tan \beta $. 
In this region the LSP annihilation fusion to
a pseudoscalar Higgs, which subsequently decays into a $b \bar{b}$ 
pair, dominates the annihilation process. 
Interestingly enough in the
same large $ \tan \beta $ region the elastic LSP - nucleon cross section
receives one of its highest possible values close to the sensitivity limits 
put by future experiments which 
will directly search for DM \cite{LNS4,Kim}.
In this region as well as in the neutralino - stau coannihilation region 
the LSP is mainly bino. 
In the focus point region the neutralino LSP 
bears a relatively sizeable higgsino component and  
its pair annihilation cross section
through the $Z$ and/or the light Higgs boson exchange becomes
significant, resulting to cosmologically acceptable relic density
and high values for the LSP-nucleon cross section. 
 
In order to calculate the LSP relic density,
one has to solve the Boltzmann equation having as input the thermal
average $\xsec$ of the LSP's annihilation 
cross section 
\cite{EHNOS,x0,x1,x2,x3,x4,x5,x6,x7,x8,x9,Belanger:2001fz,Baer:2002fv,Nihei} 
$\; \sigma (s)\;$ times their relative velocity $\;v_{rel}\;$, if we are 
in a region of the parameter space where coannihilation effects are   
negligible. 
Its computation demands integration over the center of mass scattering angle
$\; \cos \theta_{\mathrm{CM}} \;$ of the transition matrix element squared 
summed over the final spins followed by an additional integration over $s$ 
through which the aforementioned thermal average is defined.   

If one follows  the standard trace technique for calculating the 
amplitudes many interference terms are encountered since 
the squares of the annihilation amplitudes are summed 
in a coherent way. However 
this can be successfully accomplished as has been shown elsewhere 
\cite{Belanger:2001fz,Baer:2002fv,Nihei}. 

In this paper we shall pursue a partial wave analysis 
in the total angular momentum 
$\;J\;$, which can be considered as complementary to the trace technique 
and it improves 
analogous treatments existing in literature. In a partial wave expansion 
the integration over the scattering angle $\; \cos \theta_{\mathrm{CM}} \;$
 is carried out  
trivially, since the amplitudes of different $\;J\;$ are summed 
incoherently in the total cross
section avoiding numerous interference terms. This allows 
a better theoretical and numerical control of the annihilation process.  

In addition the analysis , which we employ in this paper, has the following
advantages: 
\newcounter{bean}
\begin{list}
{\roman{bean})}{\usecounter{bean}}
\item
It avoids the non--relativistic expansion resulting to 
more accurate numerical treatment of the thermally
averaged cross section especially near poles some of
which, as those of Higgs bosons, are significant
in obtaining small LSP relic densities in 
accord with recent cosmological data. 
The partial wave schemes encountered in literature usually expand the 
total cross section 
in powers of the relative velocity $\;v_{rel}\;$ keeping terms
 up to ${\cal{O}}(v_{rel}^2)$ 
{\footnote{In the first of reference \cite{Drees} such an 
expansion holds and the problems 
associated with the poles and thresholds are avoided by 
not expanding the kinematical factors 
$\beta_f$ and the s-channel propagators.  }}. 
\item
It allows contributions of partial 
waves of any $J$, where $J$ is the total angular momentum,   
offering a better approximation scheme of the partial wave 
series. The existing partial wave treatments 
are usually truncated to include terms up to p--waves in the 
orbital angular momentum $\;L\;$ 
of the initial state which are included in $\;J \leq 2 \;$. 
\item
The scheme is formulated in such a way as to allow easy handling 
of $CP$-violations residing 
in the superpotential Higgs mixing parameter  $\mu$ and the 
trilinear $A$ soft couplings 
and/or in the presence of additional supersymmetric $CP$-violating
 phases occurring in nonminimal models. 
\end{list}

The scheme we present can be directly compared to the exact 
results one obtains using 
the trace technique and from this the importance of terms larger than 
$J = 2$, omitted in other approaches, can be sought. 
As a preview, our analysis reveals that the thermal average of the  
cross sections times the relative velocity of the annihilating LSP's 
obtained using $\; J \leq 2 \;$ in the partial wave series approximates the 
exact result 
to an astonishingly good accuracy in the entire region of the constrained MSSM 
parameter space for temperatures that are relevant for the calculation 
of the relic density. We shall comment on this later on.
       
In this paper paying special emphasis to the mechanism  
for the LSP annihilation 
through the Higgs boson exchange, we
apply this refined technique in order to calculate the neutralino relic
density in the CMSSM in the presence of $CP$-violating terms 
residing in the $\mu$ and trilinear $A$ parameters 
of the CMSSM \cite{falk2} taking also into account the effect of the mixing 
occurring in the Higgs sector. The effect of the coannihilation processes, 
in regions of the parameter space where this applies, in the context of this 
partial wave approach,  will be the 
subject of a forthcoming publication. 

This paper is organized as follows:
In section 2 we give a brief outline of the helicity technique. 
In subsection 2.1  
we discuss the s- channel contributions of 
Higgs and Z - boson exchanges while in 2.2 
we present an analysis of the t and u - channel sfermion exchanges. 
In section 3 we briefly 
discuss the handling of the thermally averaged cross section  
and in section 4 we apply this technique when $CP$-violating phases are 
present. We end up  with the conclusions presented in section 5.

\section{The Helicity Formalism }

In this section we will briefly outline the helicity formalism relevant to our 
calculation of the amplitude for the annihilation of two LSP's 
to a fermion -  antifermion pair.  
 We should  evaluate the annihilation amplitudes for the above
processes distinguishing initial and final  helicity states . 
The  Dirac  spinor of a free fermion with mass $m$ , 3-momentum
$\vec{ \mathbf{p}}$ and  energy $E$ is:
\be
u( p ) = N \left(
{\begin{array}{*{20}c}
   \xi   \\
{\dfrac{{ \vec {\boldsymbol{\sigma}}  \cdot \ \vec {\mathbf {p}}  }}
{{E + m}}\xi }  \\
\end{array} }
\right)
\ee
where $\xi$ is a two-component spinor. The overall normalization 
factor 
$N = ( {\dfrac{{E + m}}
{{2m}}} )^{{1 \mathord{\left/
 {\vphantom {1 2}} \right.
 \kern-\nulldelimiterspace} 2}}$
is consistent with  $ u{\bar u}=1$.  
In order to find the spin - $\frac{1}{2}$ helicity states , we choose 
the two - component spinor $\xi$ to be eigenstates of the helicity operator 
$ \Lambda  = \frac{1}{2}\ { \vec {\boldsymbol{\sigma}}}
\cdot \hat{\mathbf {p} }$:
\be
\Lambda \; \xi _\lambda   = \lambda \xi _\lambda \qquad , 
\lambda  =  \pm \frac{1}{2} \qquad .
\ee
$\hat{\tb{p}}$ is the unit momentum vector,
$ \hat {\tb {p}}= ( {\cos \varphi \sin \theta ,\sin \varphi \sin \theta ,\cos \theta }) $, 
and the helicity eigenstates are found to be,
\be
\xi _ +   = \left( {\begin{array}{*{20}c}
   {\cos \dfrac{\theta }{2}}  \\
   {e^{i\varphi } \sin \dfrac{\theta }{2}}  \\
\end{array} } \right)
\qquad \mathrm{and} \quad
\xi _ -   = \left( {\begin{array}{*{20}c}
   { - e^{ - i\varphi } \sin \dfrac{\theta }{2}}  \\
   {\cos \dfrac{\theta }{2}}  \\
\end{array} } \right)
\ee
corresponding to spinors of helicities $+\frac{1}{2}$ and $-\frac{1}{2}$
respectively.
Therefore the one - particle helicity states for a particle,
$u( {p, \pm } )$, or antiparticle, $\upsilon( {p, \pm } )$, with helicities
$\pm \frac{1}{2}$ are given by
\be
u( {p, \pm } ) = N\left( {\begin{array}{*{20}c}
   {\xi _ \pm  }  \\
   { \pm \dfrac{{\left| {\vec {\mathbf{p}} }  \right|}}{{E + m}}\xi _ \pm  }  \\
 \end{array} } \right), \quad 
\upsilon ( {p, \pm } ) =  N \left( {\begin{array}{*{20}c}
   {\dfrac{{\left| {\vec {\mathbf {p}} } \right|}}{{E + m}}\xi _ \mp  }  \\
   { \mp \xi _ \mp  }  \\
 \end{array} } \right) \, .
\ee
Note that
$\upsilon ( {p, \pm } ) = C \bar u^T ( {p, \pm } )$.

For the  construction of a two - particle helicity state, 
one has to define appropriately the helicity spinors for the second fermion,
i.e for the fermion carrying momentum $- \vec{p}$ in the center of mass
frame. If the momentum of the first particle $ \vec{p}$ 
is pointing in the direction specified by 
the angles $ ( {\theta ,\varphi } )$, then the momentum of
the second particle is 
in the direction specified by
$ ( {\pi -\theta ,\varphi + \pi } )$.
The corresponding helicity spinors for the second fermion are,
\be
\xi _ {+}^{ '}  = \left( {\begin{array}{*{20}c}
   { -e^{-i\varphi }\sin \dfrac{\theta }{2}}  \\
   {  \cos \dfrac{\theta }{2}}  \\
\end{array} } \right)
\qquad and \qquad
\xi _{-}^{'}  = \left( {\begin{array}{*{20}c}
   { \cos \dfrac{\theta }{2}}  \\
   {e^{ i\varphi } \sin \dfrac{\theta }{2}}  \\
 \end{array} } \right)  \qquad .
\ee
To comply with the Jacob-Wick phase convention  these have been multiplied by
the appropriate phase factors $- 2 \lambda e^{ - i2\lambda \varphi }$
{\footnote{ For more details see for instance \cite{Haber} and references therein.}}. 

Any helicity amplitude can then be expressed as a sum over 
the total angular momentum $J$,
\be
\mathcal {M}_{\lambda _3 \lambda _4 ;\lambda _1 \lambda _2 }
( {s,\theta ,\varphi } ) = \sum\limits_J
{( {2J + 1} )
d_{\mu \mu '}^{( J )} ( \theta  ) \
{\mathcal{M}}_{\lambda _3 \lambda _4 ;\lambda _1 \lambda _2 }^{( J )}
( s )e^{i ( {\mu  - \mu '} )\phi } } ,
\label{partial}
\ee
where $\mu  = \lambda _1  - \lambda _2$ and
$\mu ' = \lambda _3  - \lambda _4 $. In this equation 
$\mathcal {M}_{\lambda _3 \lambda _4 ;\lambda _1
\lambda _2 }^{\left( J \right)} $
is the reduced matrix element, and the 
rotational functions $d_{\mu \mu '}^{ ( J )} ( \theta  )$
 satisfy the orthogonality relations
\be \label{eq; d_orth}
\int\limits_{ - 1}^{ + 1}
{d( {\cos \theta } ){d_{\mu \mu '}^{ ( J )}}^*(\theta )}
d_{\mu \mu '}^{( {J'} )} (\theta ) = \frac{2}{{2J + 1}} \delta _{JJ'} ,
\ee
so the interference terms are avoided when integrating the amplitudes
squared. 

\begin{figure}[t] 
\centering\includegraphics[scale=.7]{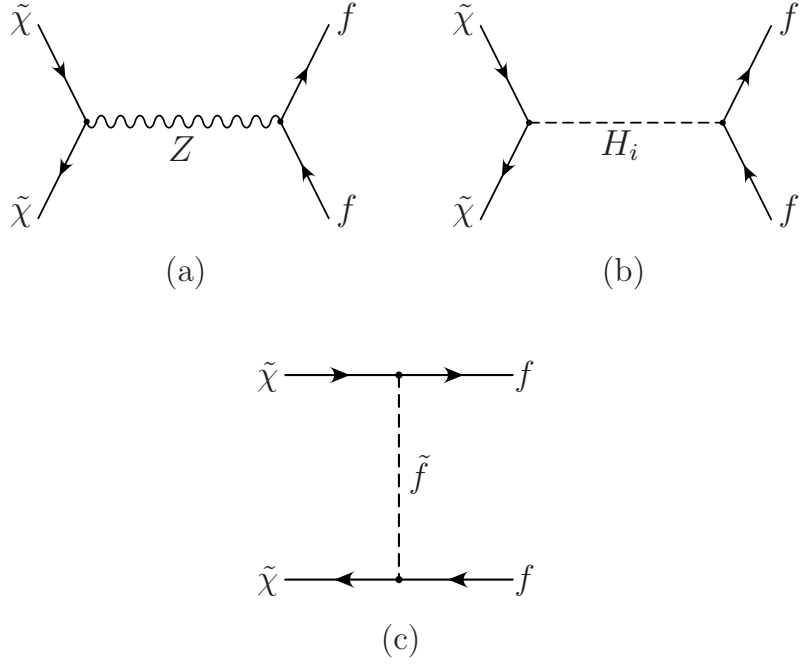}

\caption[]{Graphs that contribute to the process $\lsp \, \lsp \goes f \, \bar{f}$. 
$H_i$  in the second graph denote the Higgs mass eigenstates }
\label{feyfig:fig}  
\end{figure}

We are primarily interested in studying the annihilation of two neutralinos 
into a fermion-antifermion pair using the helicity formalism. 
The relevant channels concern $ Z , Higgs$ (s-channel)
and sfermion $\tilde f$ (t and u channel) exchange, 
see figure \ref{feyfig:fig}. 
For every helicity amplitude we need calculate  
the following matrix elements involving the spinors of the  initial, 
${  \gx {\gl 1} {p_1}, \gx {\gl 2} {p_2} }$, 
and final states, 
$ {  \fe  {\gl 3}{p_3} ,  \bar f_{\lambda _4 } ( p_4 )  }$,           
which are denoted by
$ \bra{0}
\Gamma
\ket{ \chi _{\lambda _1 } ( p_1 ) \chi _{\lambda _2 } ( p_2) }
\,
$, \hspace{4mm}
$
\bra{ f_{\lambda _3 } ( p_3 )\bar f_{\lambda _4 } ( p_4 ) }
\Gamma
\ket{0}
\,
$, \hspace{4mm}
$
\bra{f_{\lambda _3 } ( p_3 ) }
\Gamma
\ket{ \chi _{\lambda _{1,2} } ( p_{1,2}  )  }
\,
$, \hspace{4mm}
and \\
$
\bra{ \bar f_{\lambda _4 } ( p_4 )  }
\Gamma
\ket{ \chi _{\lambda _{1,2} } ( p_{1,2} )  }
\, .
$
These are the helicity dependent fundamental building blocks 
encountered in the vertices of the $s,t$ and $u-$ channel
Feynman diagrams. In these $ \Gamma $ stands for 
$ \Gamma  = \left\{ {I,\gamma ^\mu ,\gamma _5,\gamma ^\mu \gamma _5 }
\right\}$ depending on the case under consideration. 
For lack of space we do not present their analytic expressions. 
 
Each of these blocks is expressed in terms of  the helicity spinors 
described previously.
These are functions of the angle $\theta$, 
the scattering angle in the center of mass frame, assuming that the two
annihilated neutralinos move along the $z-$ axis, the azimuthal
angle $\phi $ and the center of mass energy squared  $s$. 
In the center of mass frame the magnitudes of the incoming and outgoing 
particle  3-momenta are,
\bear
p_{1,2} ( s ) = {( {\frac{s}{4} - m_{\tilde \chi }^2 } )}^{1 / 2}
\quad , \quad
p_{3,4} ( s ) = {( {\frac{s}{4} - m_f^2 } )}^{1 / 2}
\nonumber .
\eear
and the Mandelstam variables $t$ and $u$ as functions of $s$ and $\theta$
are given by,
\bear
t &=& m_{\tilde \chi }^2  +
m_f^2  - \frac{s}{2} + 2p_1 ( s )p_3 ( s )\cos \theta \quad
\nonumber\\
u &=& m_{\tilde \chi }^2  +
m_f^2  - \frac{s}{2} - 2p_1 ( s )p_3 ( s )\cos \theta  \quad .
\nonumber
\eear
In this formalism the 
unpolarized total cross section for the annihilation of 
the two neutralinos into a fermion - antifermion pair, 
in the center of mass frame, is given by 
\be
\sigma(s) \;=\; \frac{p_3}{64 \pi \; p_1 \;s } 
\sum\limits_{J}
\sum\limits_{\lambda _3 \lambda _4 ;\lambda _1 \lambda _2} \,
 (2J+1) \; 
{ |\mathcal {M}_{\lambda _3 \lambda _4 ;\lambda _1 \lambda _2 }^{(J)}|}^2  
\label{xs}
\ee
and hence no integration over the scattering angle is required 
once the helicity 
amplitudes 
${ \mathcal {M}_{\lambda _3 \lambda _4 ;\lambda _1 \lambda _2 }^{(J)}}  $
are known. Loss of accuracy occurs by truncating the 
partial wave series to a maximum 
value $J_{max}$ of the total angular momentum. 
However as stated in the introduction the 
series converges fast in the regime of energies which are relevant for the
 calculation of the 
relic density. In this paper we use a value of $J_{max}=3$ which is good 
enough to obtain accurate 
results and at the same time a fast code which returns the cross section 
and its thermal average. 
  
Using the machinery outlined before one can express any 
$s, t$ and $u - $ channel  scattering
amplitude as functions of $s$ and $\cos \theta$ and bring them 
in the form given by Eq. (\ref{partial}). 
This will provide us with the helicity amplitudes appearing in Eq. (\ref{xs}). 
The analytic expressions for the $s,t$ and $u$- channel
helicity amplitudes are listed below. 

\subsection{ s-channel contributions}
The $s$ - channel are relatively easy to handle. Since the exchanged particle
carries spin of either $s=1$, for the Z - boson, or $s=0$ for the Higgses,
the partial wave series terminates at $J \leq 1$. 

\subsubsection{ Z-exchange}
The reduced matrix element for the Z - exchange, 
as it follows from the Feynman rules using the
 helicity blocks described in the previous section,  is 
\be  \label{eq ; M_s_Z}
{\mathcal M}_{\lambda _3 \lambda _4 ;\lambda _1 \lambda _2 }^{( J ),Z}  = - \;
\frac{{ 4m_{\tilde \chi } m_f  }}
{{s - M_Z^2  + iM_Z \Gamma _Z }} \;
N_{\lambda _1 \lambda _2 }^{\mu 5}
P_{\mu \nu } \; ( {V^f F_{\lambda _3 \lambda _4 }^\nu   +
A^f F_{\lambda _3 \lambda _4 }^{\nu 5} } ) \; A_{\tilde \chi \tilde \chi } 
\ee
where the vertices
${N_{\lambda _1 \lambda _2 }^{\mu 5} } ,{F_{\lambda _3 \lambda _4 }^\nu }$
and ${F_{\lambda _3 \lambda _4 }^{\nu 5} }$ can be expressed in terms of
the following matrix elements, 
\bear
N_{\lambda _1 \lambda _2 }^{\mu 5}  &=&
\left\langle {{0\left| {\gamma ^\mu  \gamma _5 } \right.}}
\mathrel{\left | {\vphantom {{0\left| {\gamma ^\mu  \gamma _5 } \right.}
{\chi _{\lambda _1 } ( {p_1 } )\chi _{\lambda _2 } ( {p_2 } )}}}
\right. \kern-\nulldelimiterspace}
{{\chi _{\lambda _1 } ( {p_1 } )\chi _{\lambda _2 } ( {p_2 })}}
\right\rangle  \nonumber \\
F_{\lambda _3 \lambda _4 }^\nu   &=&
\left\langle {{f_{\lambda _3 } ( {p_3 } )\bar f_{\lambda _4 }
( {p_4 } )\left| {\gamma ^\mu  } \right.}}
\mathrel{\left | {\vphantom {{f_{\lambda _3 } ( {p_3 } )
\bar f_{\lambda _4 } ( {p_4 } ) \left| {\gamma ^\mu  } \right.} 0}}
\right. \kern-\nulldelimiterspace}{0} \right\rangle  \nonumber   \\
F_{\lambda _3 \lambda _4 }^{\nu 5}  &=& \left\langle {{f_{\lambda _3 }
( {p_3 } )\bar f_{\lambda _4 } ( {p_4 })\left| {\gamma ^\mu  \gamma _5 }
\right.}}
\mathrel{\left | {\vphantom {{f_{\lambda _3 } ( {p_3 } )
\bar f_{\lambda _4 } ( {p_4 } )\left| {\gamma ^\mu  \gamma _5 } \right.} 0}}
\right. \kern-\nulldelimiterspace}{0} \right\rangle  \quad . \nonumber
\eear
$ {A_{\tilde \chi \tilde \chi } } $ is the  coupling   
of the $Z-$ boson to the neutralinos and $ {V^f ,A^f } $, 
are the vector and axial couplings of the $Z-$ boson to a 
fermion - antifermion pair 
labeled by $f$.  

The numerator of the $Z$ propagator, in the unitary gauge, is 
$P_{\mu \nu }  =  - g_{\mu \nu }  + \dfrac{{k_\mu  k_\nu  }}{{M_Z^2 }}$
where $g_{\mu\nu}=diag ( {1,-1,-1,-1} )$.
Since the four-momentum of the exchanged $Z$- boson 
in the center of mass frame 
is $k_{\mu} = ( {\sqrt{s} ,\tb{0}} )$ the tensor
$P_{\mu \nu }$ receives  the following form
\bear
P_{\mu \nu }  = \left\{ {\begin{array}{*{20}c}
   & - 1 + \dfrac{s}{{M_Z^2 }}& ,&\mu  = \nu  = 0&   \nonumber \\
   &0& , &\mu  = 0,\nu  = i&   \nonumber \\
   &-g _{ij}& ,&\mu  = i, \nu  = j& . \nonumber \\
\end{array} } \right.    \nonumber
\eear
where $i,j = 1,2,3$. 

After evaluating (\ref{eq ; M_s_Z}) , we find that  a non-zero contribution
emerges only when the helicities of initial and final states are combined 
to give $J = 0  \quad \mathrm{or} \quad 1 $ as expected. In fact
\be
{\mathcal M}_{\lambda _3 \lambda _4 ;\lambda _1 \lambda _2 }^{(J = 0,1),Z}
= \frac{{ - A_{\tilde \chi \tilde \chi } }}
{{s - M_Z^2  + iM_Z \Gamma _Z }} \;
K_{\lambda _3 \lambda _4 ;\lambda _1 \lambda _2 }^{(J = 0,1),Z} .
\ee
with 
\bear
K_{\lambda _3 \lambda _4 ;\lambda _1 \lambda _2 }^{( J = 0),Z} &=& 
( { - 1} )^{\lambda _3  - \lambda _1 } 
\; \delta _{\mu 0} \delta _{\mu '0}\;
(\; 4m_{\tilde \chi } m_f \;)\;
( { - 1 + \dfrac{s}{M_Z^2 }} ) \;  A^f \nonumber \\ 
K_{\lambda _3 \lambda _4 ;\lambda _1 \lambda _2 }^{( J = 1),Z} &=& 
\frac{1}{3} ( {\delta _{\mu 1}  - \delta _{\mu  - 1} } )
\sqrt {s - 4m_{\tilde \chi }^2 } \nonumber \\
&& \left[ {\delta _{\mu '0} \;( {2m_f } )\; \sqrt 2 \; V^f  +
2 \;( {\delta _{\mu '1}
+ \delta _{\mu ' - 1} } ) \; ( {\sqrt s \; V^f  + \mu^\prime
\sqrt {s - 4m_{f }^2 }\; A^f } )} \right]
\nonumber \\
\eear 
\subsubsection{  Higgs  exchange} 
\par Using the technique outlined previously one can obtain the 
corresponding amplitudes 
for the s - channel annihilation
process through the Higgs mass eigenstates $H_i, i=1,2,3 $   
\be
{\mathcal M}_{\lambda _3 \lambda _4 
          ;\lambda _1 \lambda _2 }^{\left( J \right),H_i }  =\;-\; \frac{  1  }
{{s - M_{H_i}^2  + iM_{H_i} \Gamma _{H_i} }} \;
K_{\lambda _3 \lambda _4 ;\lambda _1 \lambda _2 }^{\left( J \right),\; H_i} 
\ee
In this equation , the superscript $H_i$ denotes the exchanged Higgs and the 
non - vanishing  
$K_{\lambda _3 \lambda _4 ;\lambda _1 \lambda _2 }^{\left( J \right),\; H_i} $ are those 
corresponding to $J=0$ only, since Higgses carry zero spin. These are given by 
\bear
K_{\lambda _3 \lambda _4 ;\lambda _1 \lambda _2 }^{( {J = 0} ),\; H_i} = \hspace{12cm} 
\nonumber \\
 \delta _{\mu 0} \delta _{\mu '0} 
\left( {g_S^{H_i} \sqrt {s - 4m_{\tilde \chi }^2 }  
+ g_A^{H_i} \sqrt s \left( { - 1} \right)^{\lambda _1  - \frac{1}{2}} } \right) 
\;
\left( {g_S^{f,i} \sqrt {s - 4m_{f }^2 }  
+ g_A^{f,i} \sqrt s \left( { - 1} \right)^{\lambda _3  + \frac{1}{2}} } \right) 
\eear
In these ,  $g_S^{H_i} , g_A^{H_i}  $ denote the scalar$(S)$ and the 
pseudoscalar $(A)$ couplings of the neutralino LSP $\tilde{\chi}$ 
to the Higgs  $H_i$, 
while $g_{S,A}^{f,i}$ are the corresponding couplings  of the same
 Higss to a fermion pair 
$f \bar{f}$,
\be
\frac{1}{2} \; \bar{\tilde{\chi}} ( g_S^{H_i} 
     + g_A^{H_i} \; \gamma_5 ) \tilde{\chi} \; H_i \;+\;
\bar{f} ( g_S^{f,i} + g_A^{f,i} \; \gamma_5 ) f \; H_i  \nonumber
\ee
Note that in our final results we should include a  
minus sign in the s - channel 
amplitudes to account for the relative (-1) difference from the corresponding  
$t$-channel amplitudes.

Note that we have allowed for the more general coupling of Higgses 
to fermions and the 
lightest of the neutralinos so that our formulae are applicable 
when supersymmetric 
$CP$-violating phases are present in the $A$ and $\mu$ parameters
 which in turn cause mixing of 
the $CP$-even and $CP$-odd Higgs eigenstates.

\subsection{t \& u-channel : sfermions exchange}
The $t$ and $u$ amplitudes are rather difficult 
to evaluate since the exchanged sfermion  
propagators depend on the scattering angle $\theta$. 
The propagators can be expanded using the formula  
\be
\frac{1}{1-a x} \;=\; \sum_{n=0} \; \Pi_n(a) P_n(x)
\ee
where $P_n(x)$ is the first 
Legendre function of order $n$ and its argument is $x=\cos \theta$. In this equation 
$a \leq 1$ and 
\be
\Pi _n ( {a } ) = \frac{{2n + 1}}
{2}\int\limits_{ - 1}^{ + 1} 
      {\frac{{P_n ( \cos\theta )}}{{1 - a \cos \theta}}d(\cos \theta)}. 
\ee
The function $\Pi_n(a)$ appearing in the above 
formula is actually  $(2n+1) \; Q_n(1/a)/a $ 
where $Q_n$ is the second Legendre function. 
Multiplying by the corresponding vertex contributions, the $t$ and $u$
channels can be cast in the form (\ref{partial}). 
Instead of presenting separately the contributions of the $t$ and $u$
channels we find it more convenient to present their sum. Note that 
a relative (-1) sign between the $t$ and $u$ channels must 
be included which has
been duly taken into account in the following expressions. We find
that their sum gives rise to the following partial wave amplitudes  
\be
{\mathcal M}_{\lambda _3 \lambda _4 ;\lambda _1 \lambda _2 }^{\left( J \right), \tilde f }  
= \sum\limits_{i=1,2} {\frac{{\left( {2J + 1} \right)^{ - 1} }}
{{F_i \left( s \right)}}\left( {\begin{array}{*{20}c}{g_S^i } & {g_A^i }  \\
\end{array} } \right)} 
\hat F_{\lambda _3 \lambda _4 ;\lambda _1 \lambda _2 }^{( J ), {\tilde f}_i } 
\left( {\begin{array}{*{20}c}
   {{g_S^i} ^* }  \\
   {{ - g_A^i} ^* }  \\
 \end{array} } \right) 
\label{amp}
\ee
$\tilde f_i$. The constants $g_S^i,g_A^i$ are 
the scalar and pseudoscalar neutralino-fermion-sfermion couplings defined as 
\be
 \bar{\tilde{\chi}} ( g_S^{i} 
    + g_A^{i} \; \gamma_5 ) \tilde{f}^{\dagger}_{i} \; f   \;+\; h.c.\nonumber
\ee
while the 
quantity $F_i( s )$ is given by 
\be
F_i \left( s \right) =  \frac{s}
{2} - \left( {m_{\tilde \chi }^2  + m_f^2  - m_{\tilde f_i }^2 } \right) \nonumber
\ee
The  matrices  
$\hat F_{\lambda _3 \lambda _4 ;\lambda _1 \lambda _2 }^{( J ), {\tilde f}_i} $
depend on the energy 
$\; E = \sqrt{s} / 2 \;$ through 
\bear
\hat A_ \pm  = \sqrt {( {E + m_f } ) ( {E + m_{\tilde \chi } } )}  
\pm \sqrt {\left( {E - m_f } \right)\left( {E - m_{\tilde \chi } } \right)} \nonumber \\
\hat B_ \pm  = \sqrt {( {E + m_f } )( {E - m_{\tilde \chi } } )}  
\pm \sqrt {\left( {E - m_f } \right)\left( {E + m_{\tilde \chi } } \right)}  \nonumber
\eear
and also on the functions $\Pi_J(a_i)$ whose arguments are  
$a_i  \equiv  \dfrac{{2p_1 p_3 }}
{{F_i ( s )}} = \dfrac{{\sqrt {( {s - 4m_f^2 } )( {s - 4m_{\tilde \chi }^2 } )} }}
{{2F_i ( s )}}$. For large energies $a_i$ approaches unity and $\Pi_J$
 approaches a logarithmic 
singularity.
In table [\ref{table11}] the matrices 
$\hat F_{\lambda _3 \lambda _4 ;\lambda _1 \lambda _2 }^{\left( J \right),\tilde f} $
are presented for all helicity combinations of the incoming LSP's 
and the outgoing 
fermions.    
\\
\begin{table}[t]
\centerline{\tb{Table 1}}
\begin{center}
\begin{tabular}[H!]{|c||c|c|} \hline 
& & \\
 $ \gl 3 \gl 4 \gl 1 \gl 2 $ &   $\mu  \mu '$ & $ 
\hat F_{\lambda _3 \lambda _4 ;\lambda _1 \lambda _2 }^{( J ), \tilde f} $  \\ 
& & \\
\hline\hline
$+ + + +$&     &$J$ even\\
$+ + -  -$ &  $\mu =0$& $
\hat M_{\lambda _1 }^{( {\lambda _3 \lambda _4 } )}  
\left[ {\Pi _J  + ( { - 1} )^{\lambda _3  - \lambda _1 } ( {\Pi _{J - 1} \dfrac{J}
{{2J - 1}} + \Pi _{J + 1} \dfrac{{J + 1}}
{{2J + 3}}} )} \right]$\\ 
$-  -  + +$& $\mu'=0$&       \\ 
$- -  -   - $&                &       \\ \hline
$+ - + +$&        & $J$ even \\
$+ -  -  -$&  $\mu=0$ & $\mu '\hat M_{\lambda _1 }^{( {\lambda _3 \lambda _4 } )}
 \left[ {\sqrt {J ( {J + 1} )} ( {\dfrac{{\Pi _{J - 1} }}
{{2J - 1}} - \dfrac{{\Pi _{J + 1} }}
{{2J + 3}}} )} \right]$ \\
$- + + +$& $\mu' \neq 0$ &              \\
$- + - - $ &                 &            \\ \hline
$+ + + -$&      & $J\geq 1$ \\
$-  -  + - $ & $\mu \neq 0,\mu '=0$ & $\dfrac{{\sqrt {J( {J + 1} )} }}{2}
\left[ { - \hat M_2^{( {\lambda _3 \lambda _4 } )}  + ( { - 1} )^J 
\hat M_3^{( {\lambda _3 \lambda _4 } )} } \right] ( {\dfrac{{\Pi _{J - 1} }}{{2J - 1}} 
- \dfrac{{\Pi _{J + 1} }}{{2J + 3}}} ) $ \\
                &    & $--------------------------$\\
$+ + - +$ &     & $( { - 1} )^J  \otimes (previous)$ \\
$-  -  - +$ &     &                  \\ \hline
$+ - + -$&      & $J\geq 1$ \\
$-  +  + - $ & $\mu \neq 0,\mu ' \neq 0$ & $
 - sign ( {\mu '} )\frac{1}
{2}\left[ { - \hat M_2^{( {\lambda _3 \lambda _4 } )}  + ( { - 1} )^J 
\hat M_3^{( {\lambda _3 \lambda _4 } )} } \right]\Pi _J  +$ \\
      &   & $+ \frac{1}{2}\left[ {\hat M_2^{( {\lambda _3 \lambda _4 } )}  
+ ( { - 1} )^J \hat M_3^{( {\lambda _3 \lambda _4 } )} } \right] 
( {\Pi _{J - 1} \dfrac{{J + 1}}
{{2J - 1}} + \Pi _{J + 1} \dfrac{J}{2J + 3}} )$ \\
&    & $--------------------------$\\
$+ - - +$ &      & $\left( { - 1} \right)^J  \otimes (previous)$ \\
$-  +  - +$ &       &        \\ \hline
\end{tabular}
\end{center}
\vspace{.2cm}
\caption{The matrices $\hat F$ appearing in Eq. (\ref{amp}). 
The matrices $\hat M$ are given in table \ref{table22}.
The $\pm$ in the helicity column  stands for $ \pm \frac{1}{2}$. }
\label{table11}
\end{table}

\begin{table}[t]
\centerline{\tb{Table 2}} 
\centerline{}
\begin{center}
\begin{tabular}[H]{|c||c| } \hline
                            &   $\hat M$ \\ \hline\hline
 $ \mu =0 ,\mu' =0 $& $
\hat M_ + ^{ +  + }  = \left( {\begin{array}{*{20}c}
   { - \hat A_ - ^2 } & { - \hat A_ -  \hat B_ -  }  \\
   { - \hat A_ -  \hat B_ -  } & { - \hat B_ - ^2 }  \\
    \end{array} } \right)$ \\
                           & $\hat M_ - ^{ +  + }  = \left( {\begin{array}{*{20}c}
   {\hat A_ + ^2 } & { - \hat A_ +  \hat B_ +  }  \\
   { - \hat A_ +  \hat B_ +  } & {\hat B_ + ^2 }  \\
  \end{array} } \right) $ \\
                         & $\hat M_ + ^{ -  - }  = \left( {\begin{array}{*{20}c}
   {\hat A_ + ^2 } & {\hat A_ +  \hat B_ +  }  \\
   {\hat A_ +  \hat B_ +  } & {\hat B_ + ^2 }  \\
    \end{array} } \right)$ \\
              &$
\hat M_ - ^{ -  - }  = \left( {\begin{array}{*{20}c}
 { - \hat A_ - ^2 } & {\hat A_ -  \hat B_ -  }  \\
 {\hat A_ -  \hat B_ -  } & { - \hat B_ - ^2 }  \\
 \end{array} } \right)$ \\ \hline
$\mu =0,\mu'\neq 0$ &$\hat M_ + ^{ +  - }  = \left( {\begin{array}{*{20}c}
{ - \hat A_ -  \hat A_ +  } & { - \hat A_ -  \hat B_ +  }  \\
{ - \hat A_ +  \hat B_ -  } & { - \hat B_ -  \hat B_ +  }  \\
\end{array} } \right)$ \\
                    &$\hat M_ - ^{ +  - }  = \left( {\begin{array}{*{20}c}
{ - \hat A_ -  \hat A_ +  } & {\hat A_ +  \hat B_ -  }  \\
{\hat A_ -  \hat B_ +  } & { - \hat B_ -  \hat B_ +  }  \\
\end{array} } \right)$ \\
                    &$\hat M_ + ^{ -  + }  = \left( {\begin{array}{*{20}c}
   {\hat A_ -  \hat A_ +  } & {\hat A_ +  \hat B_ -  }  \\
   {\hat A_ -  \hat B_ +  } & {\hat B_ -  \hat B_ +  }  \\
\end{array} } \right)$\\
                     &$\hat M_ - ^{ -  + } = \left( {\begin{array}{*{20}c}
   {\hat A_ -  \hat A_ +  } & { - \hat A_ -  \hat B_ +  }  \\
   { - \hat A_ +  \hat B_ -  } & {\hat B_ -  \hat B_ +  }  \\
\end{array} } \right)$ \\ \hline
$\mu\neq0 ,\mu'=0 $& $\hat M_2^{ +  + }  = \left( {\begin{array}{*{20}c}
   {\hat A_ -  \hat A_ +  } & { - \hat A_ -  \hat B_ +  }  \\
   {\hat A_ +  \hat B_ -  } & { - \hat B_ -  \hat B_ +  }  \\
\end{array} } \right)$ \\
                        &$\hat M_3^{ +  + }  = \left( {\begin{array}{*{20}c}
   { - \hat A_ -  \hat A_ +  } & { - \hat A_ +  \hat B_ -  }  \\
   {\hat A_ -  \hat B_ +  } & {\hat B_ -  \hat B_ +  }  \\
\end{array} } \right)$ \\
                       &$\hat M_2^{ -  - }  = \left( {\begin{array}{*{20}c}
   {\hat A_ -  \hat A_ +  } & { - \hat A_ +  \hat B_ -  }  \\
   {\hat A_ -  \hat B_ +  } & { - \hat B_ -  \hat B_ +  }  \\
\end{array} } \right)$\\
                     &$\hat M_3^{ -  - }  = \left( {\begin{array}{*{20}c}
   { - \hat A_ -  \hat A_ +  } & { - \hat A_ -  \hat B_ +  }  \\
   {\hat A_ +  \hat B_ -  } & {\hat B_ -  \hat B_ +  }  \\
\end{array} } \right)$ \\ \hline
$\mu\neq 0 ,\mu' \neq 0$ &$\hat M_2^{ +  - }  = \left( {\begin{array}{*{20}c}
   { - \hat A_ - ^2 } & {\hat A_ -  \hat B_ -  }  \\
   { - \hat A_ -  \hat B_ -  } & {\hat B_ - ^2 }  \\
\end{array} } \right)$ \\
                        &$ \hat M_3^{ +  - }  = \left( {\begin{array}{*{20}c}
   { - \hat A_ + ^2 } & { - \hat A_ +  \hat B_ +  }  \\
   {\hat A_ +  \hat B_ +  } & {\hat B_ + ^2 }  \\
\end{array} } \right)$\\
                        &$\hat M_2^{ -  + }  = \left( {\begin{array}{*{20}c}
   { - \hat A_ + ^2 } & {\hat A_ +  \hat B_ +  }  \\
   { - \hat A_ +  \hat B_ +  } & {\hat B_ + ^2 }  \\
\end{array} } \right)$\\
                         &$\hat M_3^{ -  + }  = \left( {\begin{array}{*{20}c}
   { - \hat A_ - ^2 } & { - \hat A_ -  \hat B_ -  }  \\
   {\hat A_ -  \hat B_ -  } & {\hat B_ - ^2 }  \\
\end{array} } \right)$\\ \hline
\end{tabular}
\end{center}
\vspace{.4cm}
\caption{The matrices $\hat M$ appearing in table {\ref{table11}}.}
\label{table22}
\end{table}
\section{Handling the thermal average and the relic density }

The primarily task in calculating the relic density is to find 
the thermal integral {\footnote{The $W(s)$ that used here
is related to the cross section by $\sigma(s) = W(s)/  p \sqrt{s}  $
(see Ref.~\cite{x1}).} }
\beq
\left\langle \sigma v_{rel} \right\rangle(T) = \frac{1}{2 \, \mlsp^4 \, T} \,
\frac{1}{(K_2(\mlsp/T))^2} \,
      \int_{4 \mlsp^2}^{\infty} ds \, p \, W(s) \,K_1(\sqrt{s}/T) \, .
\label{eq:thave}
\eeq 
The momentum $p$ is
related to the CM energy and $\mlsp$ through $s=4\,(p^2+\mlsp^2)$.
Scaling the temperature in terms of $\mlsp$ 
we can use as thermal variable $x=\frac{T}{\mlsp}$. 
Introducing a new variable $y$ defined by
\beq
y=\frac{1}{x} \left( \frac{\sqrt{s}}{\mlsp} -2 \right) \, ,
\label{eq:ydef}
\eeq
Eq.~(\ref{eq:thave}) can be cast in the form
\beq
\left\langle \sigma v_{rel} \right\rangle(x) = \frac{1}{2 \, \mlsp^2} \,
\frac{1}{(K_2(1/x))^2} \,
  \int_{0}^{\infty} dy \, (xy+2) \, \sqrt{xy(xy+4)} 
           \, W(y) \, K_1(y+2/x) \, .
\label{eq:thavey}
\eeq 
In this equation $W(y)$ is defined to be $W(s)$ with $s$ replaced
by its value from Eq.~(\ref{eq:ydef}), that is $\sqrt{s}= {\mlsp} (x y+2)$. 
Before continuing we should perhaps comment on how the singular behavior 
of the Bessel functions 
encountered in (\ref{eq:thave}), or same (\ref{eq:thavey}), are evaded in our 
numerical code. This problem 
is not related to the partial wave expansion we employ  to calculate the 
cross sections but it is due solely to the behavior of the Bessel functions 
appearing outside and within the integrands in the above relations. 
In fact the presence of the $K_2(1/x)$ in the
denominator of Eq. (\ref{eq:thave}) is potentially  dangerous, 
since for $ x < x_c $, with $x_c$ a small number of order 0.003 or so, 
the Bessel function underflows $K_2 \approx 0$, 
resulting to numerical overflow. 
However, the asymptotic expansion of the Bessel functions for
large arguments is known to be
\beq
K_n(z) \underset{z>c}{\too}  \sqrt{\frac{\pi}{2\,z}} \, e^{-z} \, P_n(z) \,,
\label{eq:besex}
\eeq
where $c = {1}/{x_c}$ and 
\beq
P_n(z)=1+ \frac{4n^2-1}{1!\,8z} 
     + \frac{(4n^2-1)(4n^2-3^2)}{2!\,(8z)^2} + \dots
\eeq
Using Eq.~(\ref{eq:besex}) one can obtain a form of Eq.~(\ref{eq:thavey})
which is free of such overflows in the offending region $ x <  x_c$
\beq
\left\langle \sigma v_{rel}\right\rangle(x) \underset{x < x_c}{\too} 
   \frac{1}{ \mlsp^2} \,\sqrt{\frac{1}{2 \; \pi}}\,   \frac{1}{P_2(1/x)^2} \,
  \int_{0}^{\infty} dy \, \sqrt{y(xy+2)(xy+4)} 
           \, W(y) \,  e^{-y} \,\, P_1(y+2/x) \, .
\label{eq:thavex}
\eeq 
For $ x > x_c $ the $K_2$ does not underflow but the Bessel function $K_1$
within
the integral does in the integration region in which its argument exceeds
$1 / x_c$.
In this integration region and in order to avoid such a numerical underflow
we  approximate $K_1$ by  Eq.~(\ref{eq:besex}) within the integral. 
These results can be applied for an accurate numerical evaluation
of the integral of Eq.~(\ref{eq:thavey}) valid for any $x$. 

\begin{figure}[t] 
\begin{center}
\includegraphics[width=10cm]{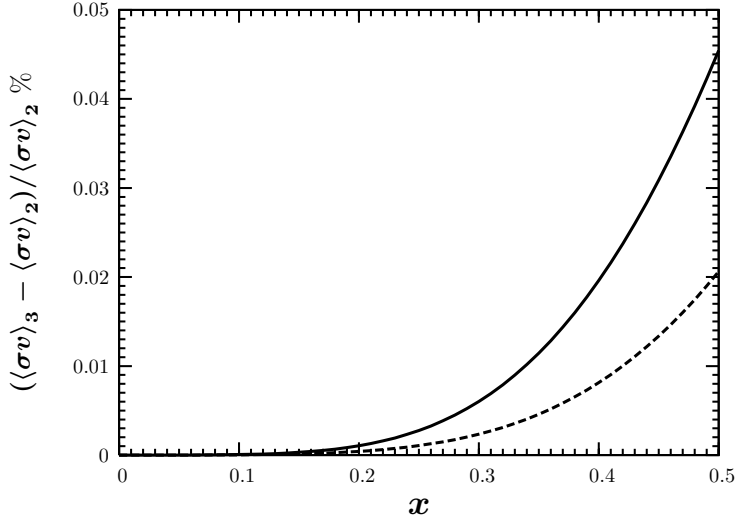}
\end{center}

\caption[]{The ratio 
$\; ({\left\langle \sigma v_{rel} \right\rangle}_{3}\;-\;
{\left\langle \sigma v_{rel} \right\rangle}_{2})/
{\left\langle \sigma v_{rel} \right\rangle}_{2}$, in $\%$, as
a function of $x$.
The subscripts refer to the value of $J_{max}$ in each case. 
The figure corresponds to 
values $A_0=0, \; m_0=1.2 \TeV, \; \tan \beta=55$ 
and values of $m_{1/2}$ equal to $\;900 \GeV$ (solid line) and 
$\;800 \GeV$ (dashed line).}
\label{fig:error}  
\end{figure}
In obtaining the LSP relic density one has to solve the 
Boltzmann transport equation 
numerically or use an approximate solution 
which however is quite accurate. Regarding the first 
approach 
the details of how one solves the Boltzmann equation are described             
by Edsjo and Gondolo and by Gelmini and Gondolo in reference \cite{x9}. 
Also an efficient method, reminiscent of the WKB approximation, 
for solving Boltzmann equation is presented in \cite{x5}. 
The details concerning the second approach, which 
is mostly used in the literature, are given  
in \cite{EHNOS, Kolb}, or by Griest et al. 
in the second of references in \cite{x3}.  
According to it the freeze - out temperature $T_f$ and the 
corresponding point $x_f$ is determined by 
\beq
x_f\;=\;\left[ \; ln \left(   \frac{0.038 \; g_{eff}\; M_{Planck} \; m_{\lsp}  
\left\langle \sigma v_{rel} \right\rangle}{g_{*}^{1/2}}\; x_f^{1/2}
\right) \right]^{-1}  \; \; .  \label{freeze}
\eeq
In this equation $g_{eff}$ are the effective
energy degrees of freedom and $g_{*}^{1/2}$ 
is the total number of effective relativistic degrees of freedom related to 
the entropy degrees of freedom $h$ and $g_{eff}$  by  
\beq
g_{*}^{1/2}\;=\; \left( \; h + \frac{m_{\lsp}}{3} \; \frac{dh}{dT} \;\right)\;
\frac{1}{\sqrt{g_{eff}}} \; \; .
\eeq
Eq. (\ref{freeze}) can be iteratively solved to yield the freeze - out  
value $x_f$ and 
from this by integrating  $ \left\langle \sigma v_{rel} \right\rangle(x) $ 
one can have the integral
\beq
J_{*} \;=\; \int_{0}^{x_f} \; \left\langle \sigma v_{rel} \right\rangle(x) \; dx \; .
\label{jjj}
\eeq
The LSP relic density is then given by the expression 
\beq
\Omega_{\lsp} h^2 \;=\; \frac{1.07 \times 10^{9} \; GeV^{-1}}
{g_{*}^{1/2}\; M_{Planck}\;J_{*} } \; \; . \label{omom}
\eeq
The value of the freeze - out point $x_f$ is 
roughly $\; \approx 1/20 \;$ and thus the values of 
$\left\langle \sigma v_{rel} \right\rangle$ needed 
for for the calculation of  (\ref{omom}) refer to points  
$\; x \leq x_f \;$ as is obvious from (\ref{jjj}). 

In order to estimate the error induced by 
truncating the partial wave series to values $J \leq J_{max}$ 
in figure \ref{fig:error} we 
display the ratio 
$\; ({\left\langle \sigma v_{rel} \right\rangle}_{3}\;-\;
{\left\langle \sigma v_{rel} \right\rangle}_{2})/
{\left\langle \sigma v_{rel} \right\rangle}_{2}$, 
where the subscripts refer to the value of $J_{max}$ in each case. 
The plot of the figure corresponds to 
values $A_0=0,\; m_0=1.2 \TeV,\; \tan \beta=55$ and values of $ m_{1/2}$ 
equal to $ 900\GeV$ (solid line) and $800\GeV$ (dashed line). Both 
cases refer to $\; \mu > 0 $. 
The figure clearly shows that this ratio is small 
$\; \simeq 10^{-4} \;$ when $x$ lies 
between $0$ and $0.5$, which includes the
range of integration in Eq.~\ref{jjj}, indicating that
the partial wave series saturates the exact result
 if one truncates at  a value $J_{max}=2$. 
This is a general feature valid in the entire 
region of the parameter space. In our numerical
 analyses concerning this work  the value of $J_{max}$,
 which is actually input in our code, is set to  $J_{max}=3$. 
Higher values will yield more precise results at the expense 
of slowing down our numerical code. 

It should be pointed out that the partial wave expansion we employ 
in terms of the total angular momentum $J$ differs from the expansion
 in terms of the orbital angular momentum $L$. Since      $\; L+S \;$ 
is even for the initial state of the two Majorana LSPs, expanding up to 
$J=2$ includes not only the $L=1$ terms (p--waves) but in addition all 
amplitudes characterized by $L=2$. This is much improvement over the 
existing orbital angular momentum schemes which are truncated to $L=1$ 
equivalent to keeping terms up to ${\cal{O}}(v_{rel}^2)$ 
in the relative velocity of the two LSPs. For this  reason  the series in
 terms of $J$ yields very accurate results if already truncated at 
$J_{max}=2$, as shown in figure \ref{fig:error}.

\section{Neutralino abundance with $CP$-violation at large $\tanb$}
 We apply the above described helicity amplitude technique to 
calculate the neutralino relic density at the large $\tanb$ regime, where
it is known that in the $CP$-conserving case  the contribution for 
the neutralino pair annihilation
cross-section through the pseudo-scalar Higgs boson $A$ exchange 
plays an important role in obtaining cosmologically acceptable 
relic densities far from the coannihilation and focus point regions.  
Since in this work we are mainly interested in the aforementioned 
mechanism for the annihilation of the LSP's, other 
production channels, e.g. gauge, Higgs bosons etc, 
do not contribute significantly. 
Hence we consider the neutralino LSP pair annihilation to fermion 
pairs at  the relativistic level, including all finite width contributions. 
The code we employ,  
which is based on the partial wave analysis 
we outlined in the previous chapters, is pretty fast and 
can cope with the Higgs 
resonances which play an important role in obtaining an LSP 
relic abundance in accord with the recent cosmological data.
As already stated in the previous section the value of the 
maximum allowed angular momentum 
$J_{max}$ is input but a value of $J_{max}=3$ is sufficient in obtaining a 
very good accuracy.

\begin{figure}[t] 
\begin{center}
\includegraphics[width=7cm]{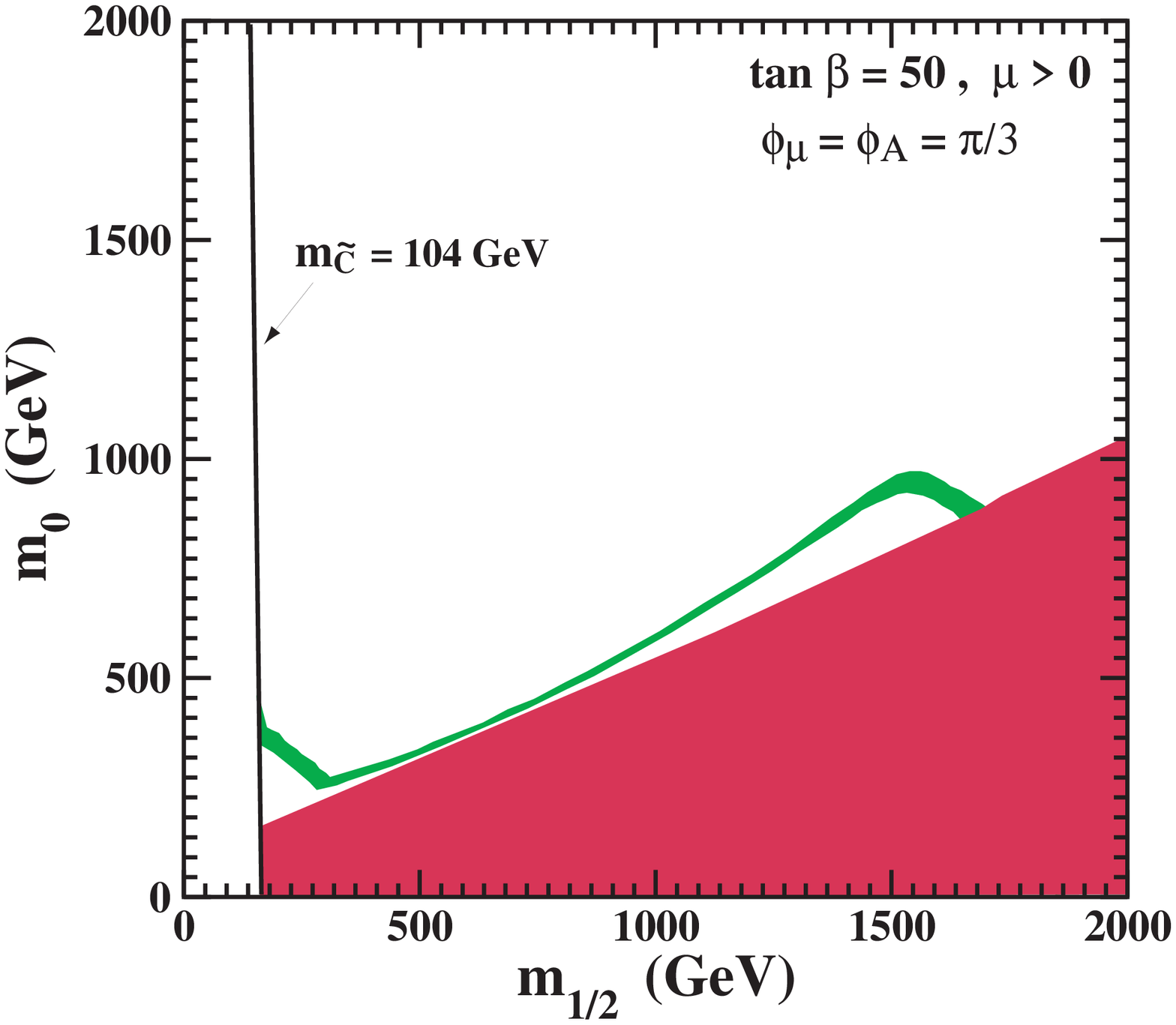}
\includegraphics[width=7cm]{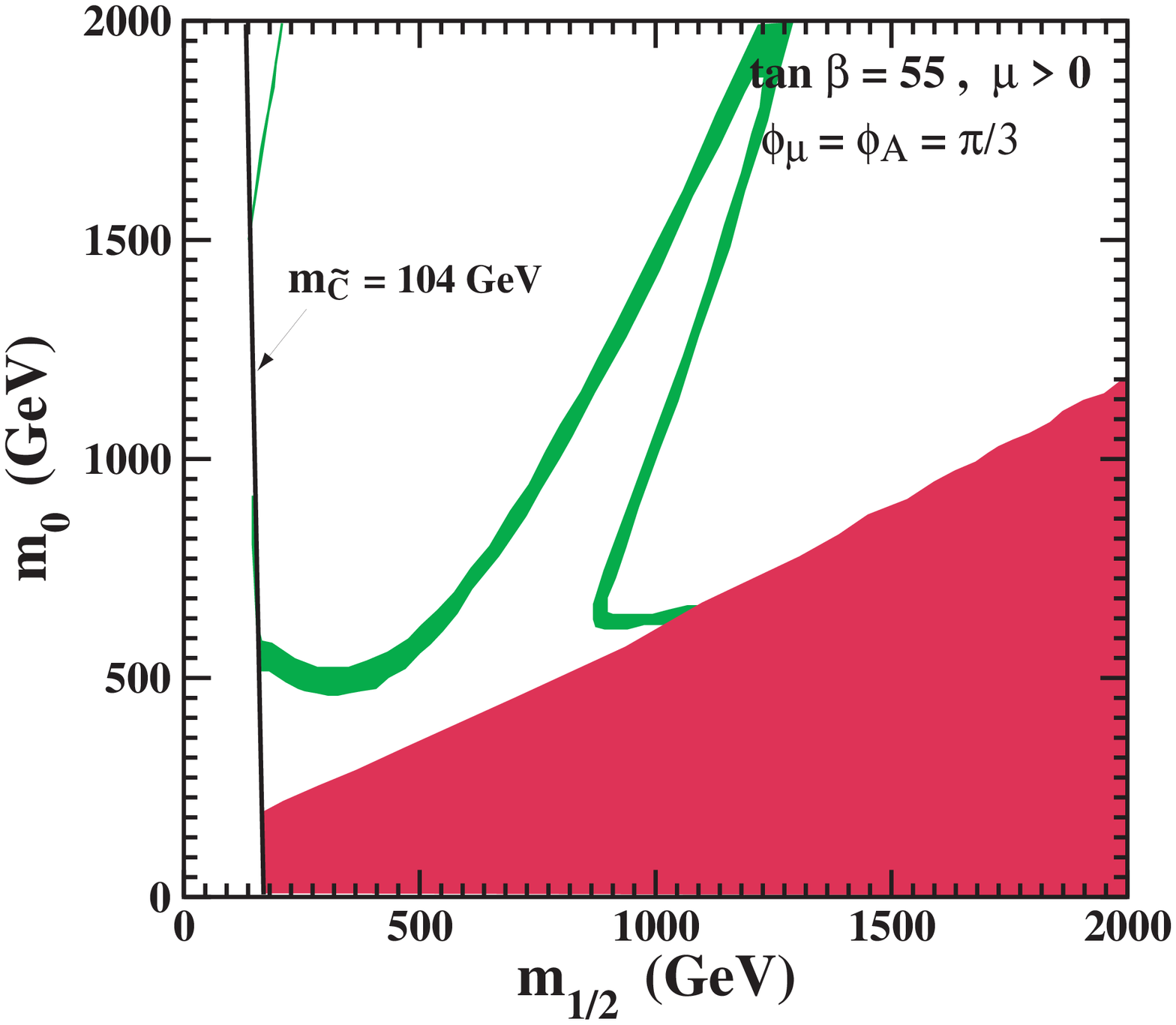}
\end{center}

\caption[]{The neutralino relic density, in the $(m_0,m_{1/2})$ plane,
 for $\tan\beta=50$ and $55$. }
\label{feyfig:planes}  
\end{figure}

\begin{figure}[t] 
\begin{center}
\includegraphics[width=7cm]{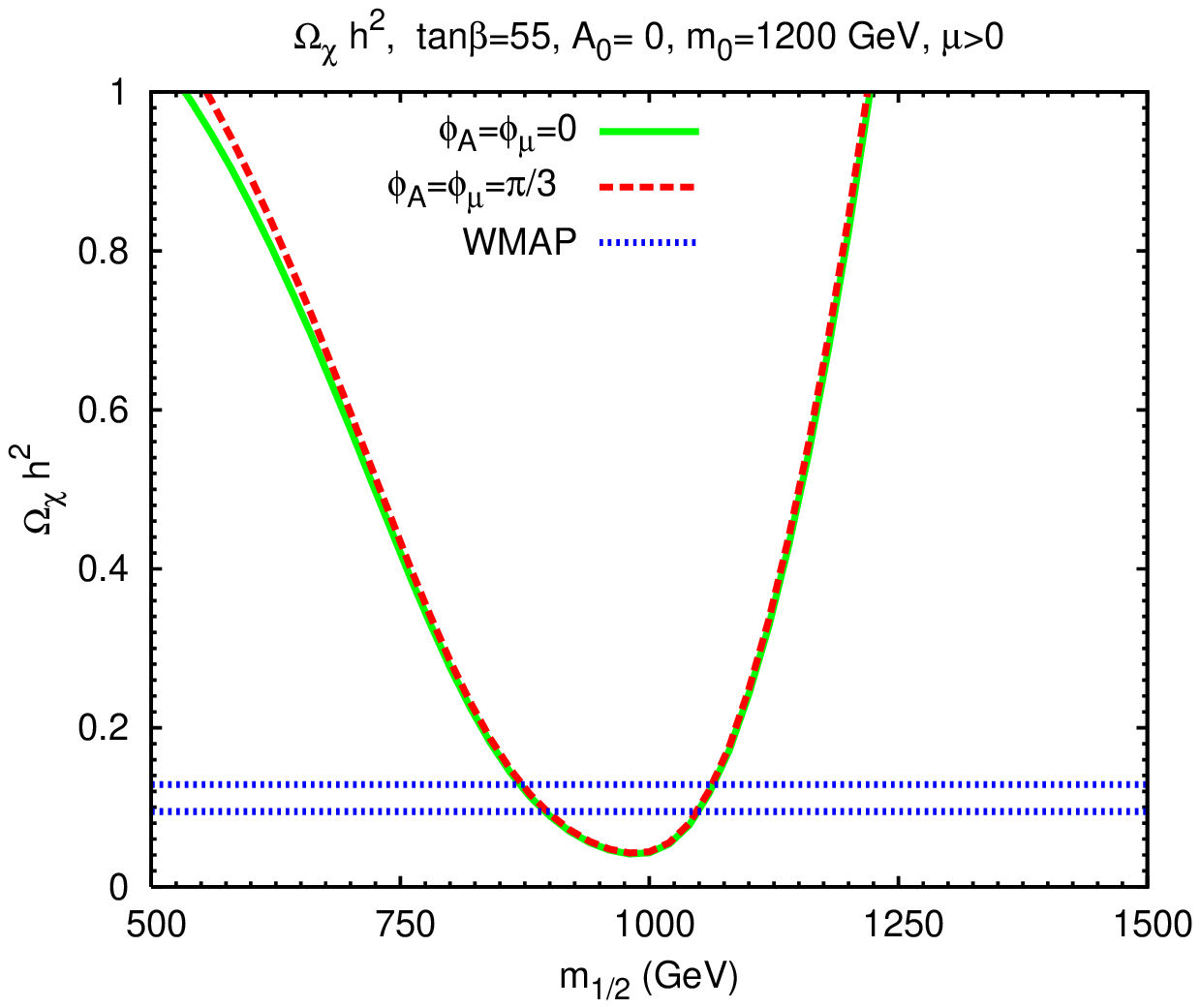}
\includegraphics[width=7cm]{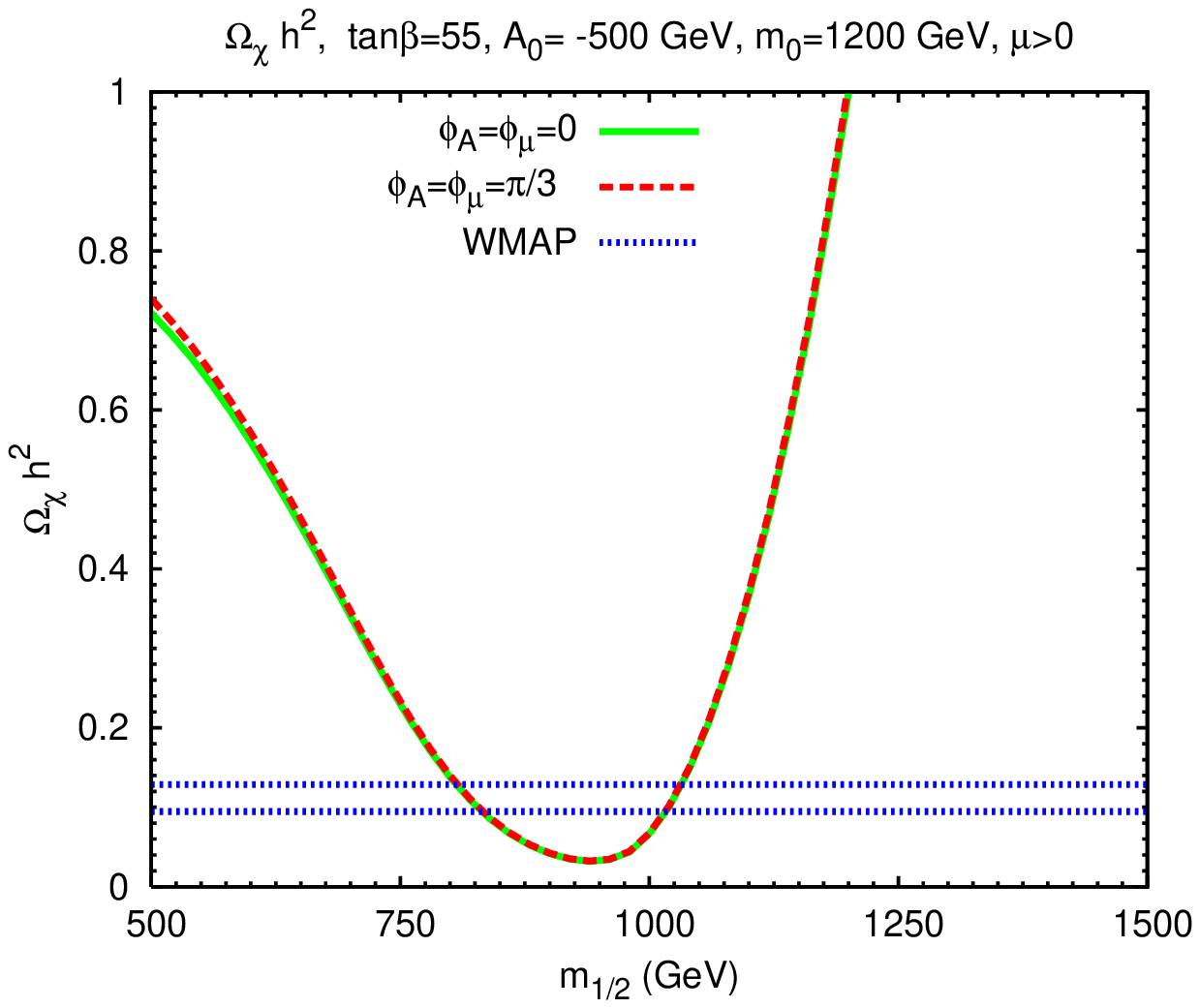}
\end{center}
\caption[]{The relic density $\relic$ for fixed $m_0$, 
as a function of $m_{1/2}$ 
for  values of the phases and the remaining parameters 
shown in the figure. Only the $\mu>0$ cases are shown. 
For comparison the corresponding values when the phases are
 zero are displayed (solid line). 
The horizontal dotted lines mark the cosmologically allowed region from 
the WMAP data.}
\label{feyfig:cuts}  
\end{figure}

In this work we have focused our interest in the so-called 
``funnel'' region of the parameter space, where one obtains cosmological
accepted relic density through annihilation to 
Higgs boson resonances. In principle 
one expects that
relatively sizable $CP$-violating effect can also occur  
in the focus point region,
where the LSP carries a non-zero Higgsino component.  This results to
an enhancement of the LSP pair annihilation  cross section
to massless fermions, through the $Z$-boson exchange. Although we will not
present numerical results in the  focus point region, 
we have checked that there
the $CP$-violating effects  are almost comparable with those around
the Higgs pole. Furthermore, the latest  
CDF and D0 data~\cite{mtop}
suggest a somewhat higher value for the top quark mass
 $m_t = 178$~GeV, pushing the focus point region up to considerably high
values of $m_0$.

The method we develop is capable of accommodating  
 $CP$-violating sources residing in the $\mu$ 
and the trilinear $A$ parameters as well as in other parameters, 
as for instance in the gaugino mass parameters, which are important 
when we depart  
from the minimal schemes. In this work we consider the 
constrained scenario and hence only the $\mu $ 
and $A$ phases are allowed. In figure \ref{feyfig:planes} 
and for nonvanishing phases equal to $\phi_\mu, \phi_A = \pi/3$ 
 at the weak scale 
we present in the $m_0, m_{1/2}$ plane the allowed domains (in green),
 which are consistent with the recent cosmological data  
from WMAP \cite{wmap} $\;\relic = 0.1126^{+ 0.0161}_{-0.0181} \;$. 
The cases shown are for values of $\tan \beta$ equal to $50$ and $55$ 
respectively. The shaded area (in magenta) at the bottom, in both panels, 
is excluded 
since in this the stau is the LSP. In figure \ref{feyfig:cuts} we 
display the relic density for fixed $m_0, A_0$ as a function of $m_{1/2}$ 
for $\tan \beta = 55$ and for nonvanishing phases $\phi_\mu, \phi_A = \pi/3$  
of $\mu$ and  $A$. For comparison the case with  $\phi_\mu, \phi_A = 0$ 
is also shown.  
In the left and right panel of this figure the cases  $A_0=0$ and 
$A_0=-500 \GeV$ are presented.
We should remark that in our analysis we have relaxed the 
EDM constraints \cite{edm,deuteron} 
in order to locate possible regions of the parameter space which yield  
substantial deviations from the results of the $CP$-conserving case.

From figure \ref{feyfig:cuts} one sees that no significant change is 
observed and the
results for the $CP$-conserving and $CP$-violating cases are almost 
identical. This holds for other values of $m_0, A_0$ and phases as well 
showing that the effect of the phases plays no important role to the 
relic density. Phrased in another way, despite the fact that 
the $CP$-even and $CP$-odd Higgs eigenstates get mixed when $CP$ 
is violated, the Higgs exchange is still the dominant mechanism 
in obtaining small values of the relic density, as long as we are away 
from the coannihilation and focus point regions. 
The mixing of Higgses \cite{pilaf2,ibrah} induced by the phases
 of $\mu$ and $A $
has little effect and this is in agreement with 
the findings of Ref.~\cite{Nath}.   
Thus the $CP$-conserving 
case results are not destabilized by switching on the 
phases of $\phi_\mu, \phi_A $ parameters and the 
conclusions reached in previous analyses \cite{LNS1,LNS4,LS,LN} hold 
true. 

\begin{figure}[t] 
\begin{center}
\includegraphics[width=7cm]{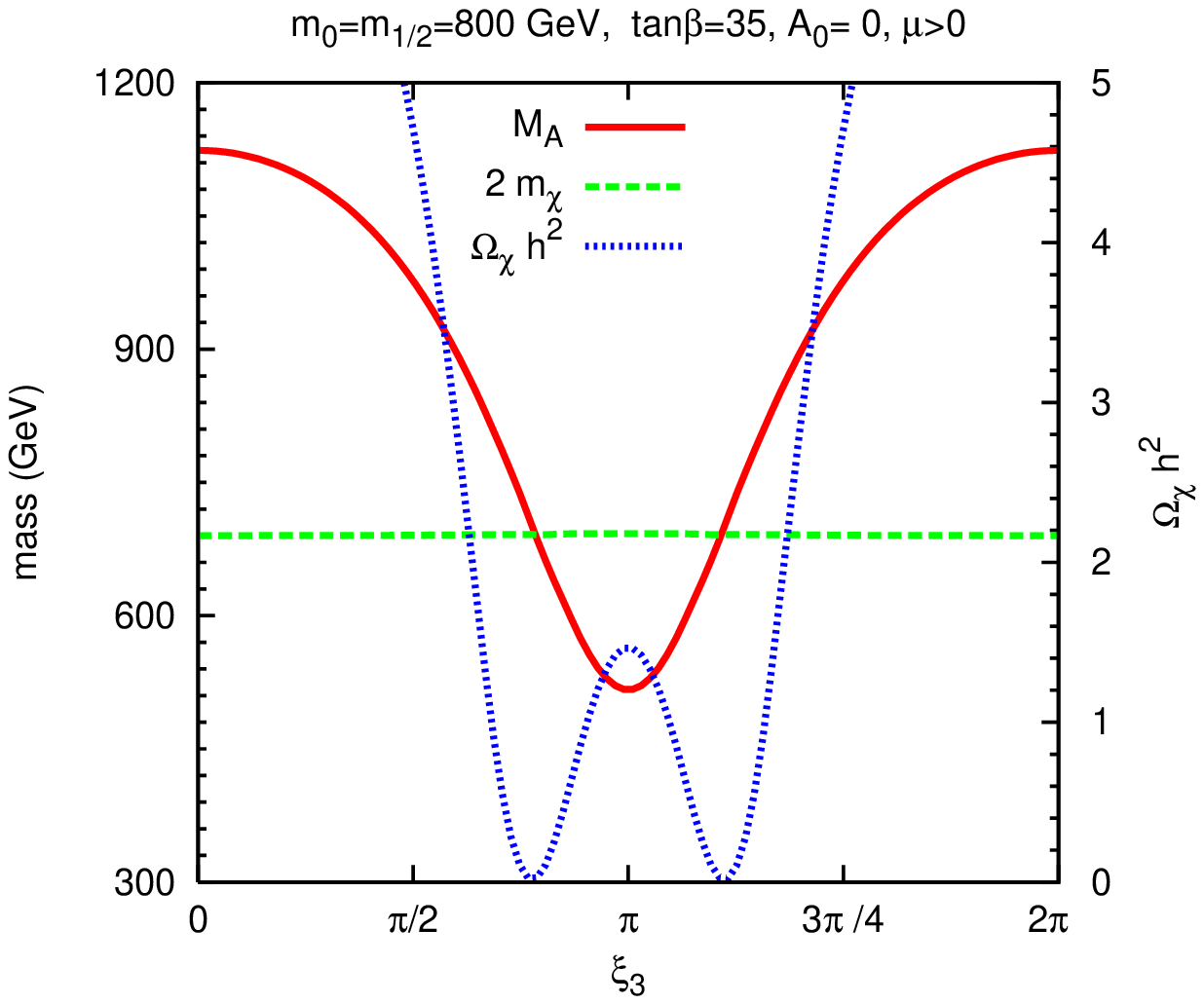}
\includegraphics[width=7cm]{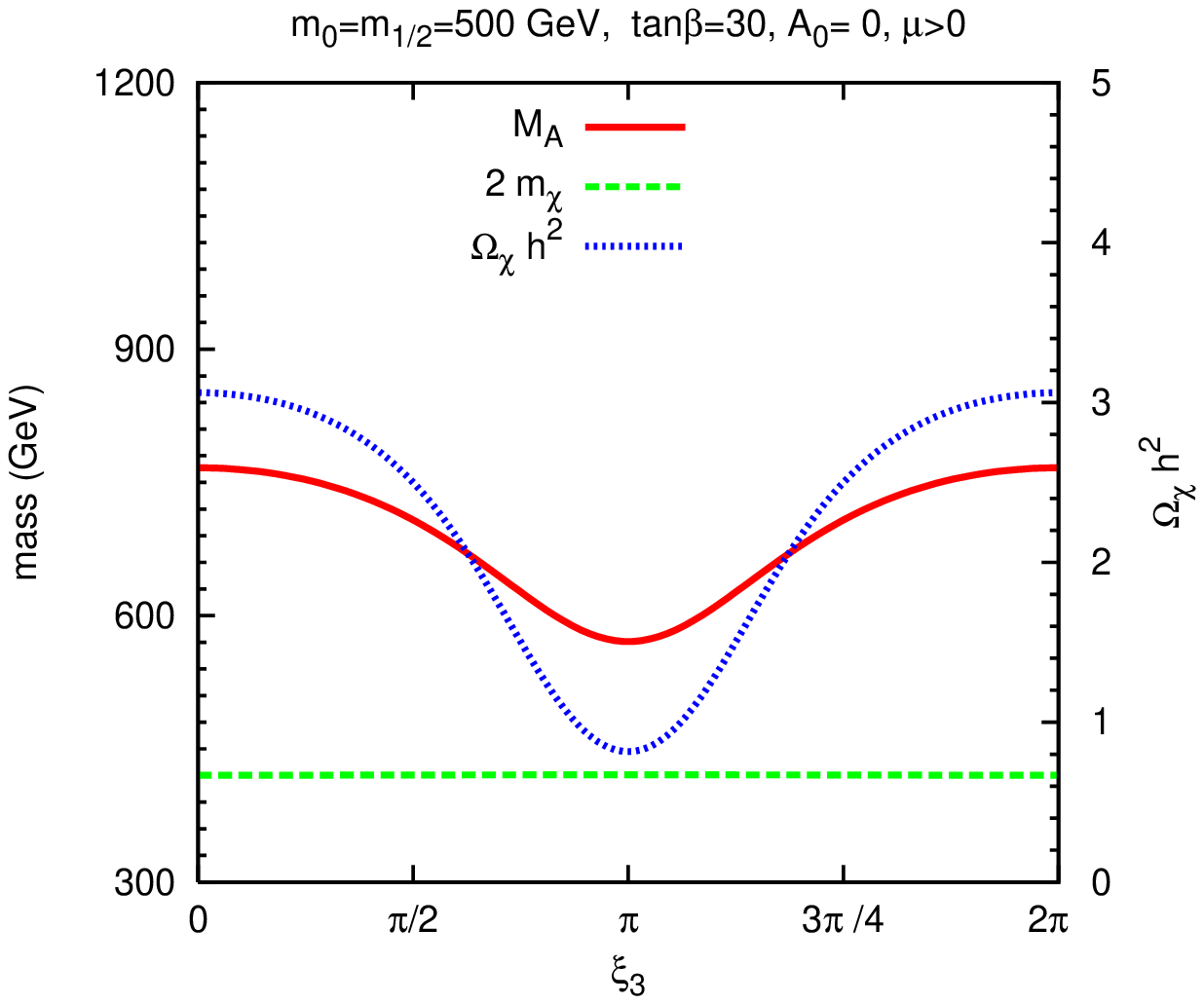}
\end{center}

\caption[]{The mass $M_A$ (solid line), $\mlsp$ (dashed line) 
and the neutralino relic density (dotted line) as functions of $\xi_3$ 
when all other phases are set to zero. In the left (right)
panel  $m_0=m_{1/2}=800 \GeV, \tanb=35$ ($m_0=m_{1/2}=500 \GeV,\tanb=30$), 
and  $ A_0=0, \mu>0 $.  }
\label{fig:x3cut1}  
\end{figure}

Significant changes in the relic density can occur if one allows for 
the existence of phases in the gaugino mass parameters mainly 
because the latter  
affect the value of the bottom quark mass as shown in \cite{Nath}.
In particular, the supersymmetric QCD corrections to the bottom
quark mass depend on the phase $\xi_3$ of the gluino mass $M_3$ 
at the GUT scale. 
Even if we ignore the phases $\phi_\mu, \phi_A $ as well as the 
phases of the gaugino masses $M_1,M_2$ and the angle specifying 
the misalignment of the vev's of the Higgs fields, the effect
 of $\xi_3$ is quite important and can drastically change the 
picture. In the discussion that follows  we have only allowed 
for a nonvanishing $\xi_3$  setting all other phases to zero.
 With only $\xi_3$ present the electroweak sector is not 
directly affected by it, up to the one loop order. 
At this order the phase $\xi_3$ affects the electroweak
 sector implicitly  through changes in the bottom Yukawa 
coupling as described earlier. In particular the Higgs 
sector is the same as in the $CP$-conserving case and the 
1-loop expressions for Higgs masses still hold.  
It is known that the mass of the pseudoscalar 
Higgs $A$ is quite sensitive to the
size of low energy threshold supersymmetric corrections to the
bottom Yukawa coupling~\cite{LS,Nath}.  Consequently $\xi_3$ influences 
the position of the $A$-Higgs pole,  $M_A\simeq 2\mlsp$, where
the rapid  neutralino pair annihilation yields cosmologically acceptable
relic densities. In order to see 
the effect of the phase of $M_3$ on the relic density, in 
fig.~\ref{fig:x3cut1} we plot the values of the pseudoscalar mass $M_A$ , 
the value of the mass of the annihilated neutralino pair $2 m_{\tilde \chi}$ 
and the relic density as functions of the phase $\xi_3$ for two 
characteristic sets of input parameters shown on the left and 
right panel. As  already stated we have only switched  on 
the phase $\xi_3$. On the left panel it is observed that 
although the value of $\tan \beta$ is not quite high the 
approach to the pole, $M_A\simeq 2\mlsp$, takes place 
for values $\xi_3\simeq 3\pi/4$ resulting 
to rapid $\lsp$  annihilation through $A$-Higgs boson 
exchange and small relic densities in accord the WMAP 
data ($\relic \simeq 0.1$). On the right panel and for 
different input values we observe that the relic density 
becomes significantly smaller for values $\xi\simeq \pi$, but not compatible
with the WMAP bound. In both case the plots look symmetric upon 
$\xi_3 \goes -\xi_3$. This is due to the fact that the SQCD corrections
 are insensitive to the sign of $\xi_3$ if the other phases are put to zero. 

\begin{figure}[t] 
\begin{center}
\includegraphics[height=6.6cm,width=7cm]{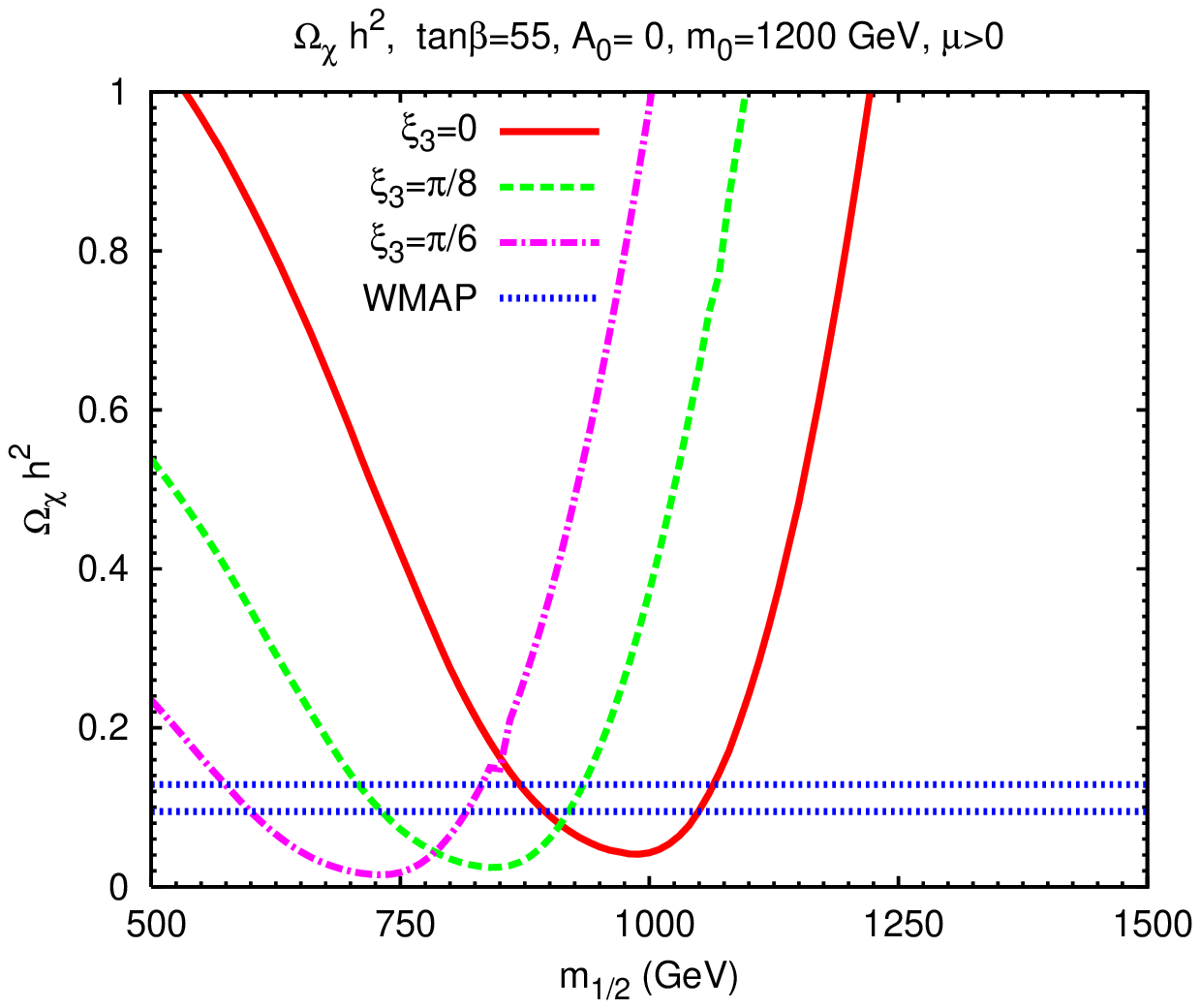}
\includegraphics[width=7cm]{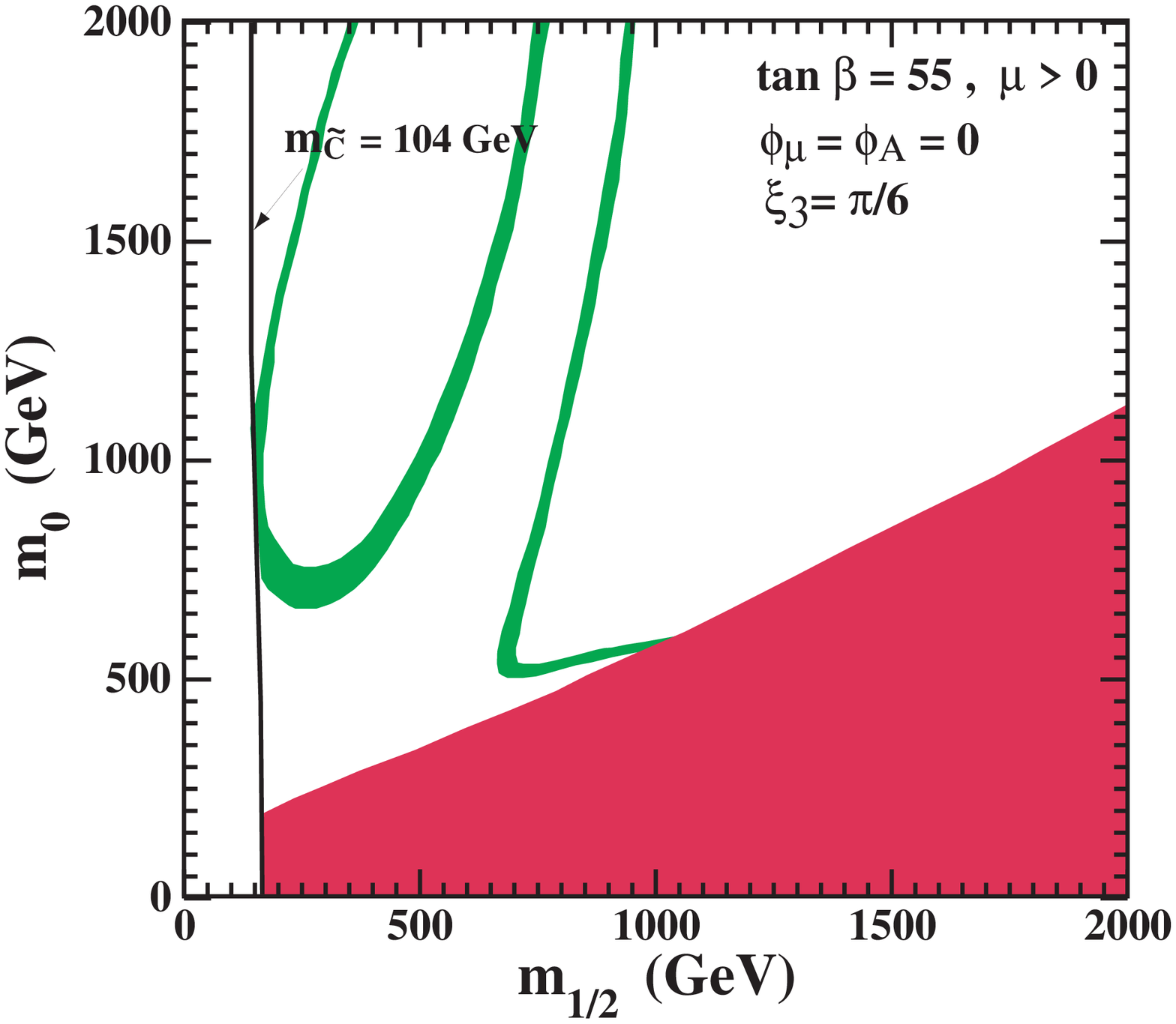}
\end{center}
\caption[]{(Left panel) The neutralino relic density 
as a function of $m_{1/2}$,
for $m_0=1200 \GeV, A_0=0, \tanb=55, \mu>0$, 
for $\xi_3=0, \; \pi/8$ and $ \pi/6$ 
(solid, dashed and dashed-dotted lines). The WMAP favored region is marked
 by the horizontal dotted lines. \\
(Right panel) The neutralino relic density, in the $(m_0,m_{1/2})$ plane,
 for $\tan\beta=55$  and $\xi_3$ equal to $\pi /6$. }
\label{fig:x3cut2}  
\end{figure}

In  the left panel of figure ~\ref{fig:x3cut2} 
we plot the neutralino relic density
for various values of $\xi_3$ and the same input parameters 
$m_0=1200 \GeV, A_0=0, \tanb=55, \mu>0$.
One can see that for $\xi_3=0$, $CP$-conserving case, 
the minimum of the $\relic$ appears at
$m_{1/2}=1000 \GeV$, while for $\xi_3=\pi/6$  
this occurs at a lower value $m_{1/2}=750 \GeV$. 
This means  that as $\xi_3$ increases,  
the funnel of the right panel of figure ~\ref{feyfig:planes} 
is widening up and moving to the 
left allowing for smaller values of $m_{1/2}$ and hence 
lighter supersymmetric mass 
spectrum in general. This is clearly seen in the 
right panel of figure ~\ref{fig:x3cut2} 
which should be compared to the right panel of figure ~\ref{feyfig:planes} 
{\footnote{In figure ~\ref{feyfig:planes} the phases $\phi_{\mu},\phi_{A}$ have
 been taken equal to 
$\pi/3$ and not vanishing as in figure ~\ref{fig:x3cut2}.
 However when $\xi_3$ is 
zero the effect of $\phi_{\mu},\phi_{A}$ is negligible and the 
figure ~\ref{feyfig:planes} 
is the same with the corresponding one 
in which $\phi_{\mu},\phi_{A}$ are put to zero.   }}.

\section{Conclusions and Outlook}
We have presented an improved partial wave expansion technique for
calculating the neutralino pair annihilation to a fermion pair, based
on the helicity amplitude method. In the partial wave treatment  
the amplitudes carrying different total angular momenta  
are summed incoherently in the total cross
section, avoiding numerous interference terms, resulting to  
a better theoretical and numerical control of the neutralino 
annihilation process.  
Also the non  relativistic expansion is avoided obtaining a  
more accurate numerical treatment of the thermally
averaged cross section especially near poles some of
which, as those of Higgses, are significant
in obtaining small LSP relic densities as recent cosmological data suggest. 
The scheme we employ allows for contributions of partial waves of any $J$,  
offering a better approximation of the partial wave series. The  
methods existing in the literature  are usually truncated to include terms
 up to p - waves 
in the orbital angular momentum $L$ of the initial state of the annihilating  
LSP's corresponding to values of $J \leq 2$. 
Our scheme is formulated in such a way as to allow easy handling of 
$CP$-violations residing in the superpotential Higgs mixing parameter  
$\mu$ and the trilinear $A$ soft couplings 
as well as  in the presence of additional $CP$-violating 
phases occurring in nonminimal 
extensions of the MSSM.  

Using this method we include non - vanishing $CP$-violating phases  
$\phi_\mu, \phi_A $ to the $\mu$ and $A$ parameters of the CMSSM. 
We study the effect of these phases at the 
large $\tanb$ regime, where it is known that the exchanges of Higgses is
 one of the dominant mechanisms to reduce the values of the LSP relic
 density to cosmologically acceptable levels 
if $CP$ is conserved.  
We found that even if one relaxes the EDM constraints 
the effect of these phases 
is small, in the entire region of the parameter space away 
from the coannihilation 
and focus point regions, and the  
values of the neutralino relic density obtained are almost the same as in the 
$CP$-conserving case. 
Therefore the Higgs exchange mechanism is still one of the dominant 
mechanism in obtaining small values for the relic density and 
the $CP$-conserving case results are not upset by switching on the
 phases  $\phi_\mu, \phi_A $. Significant changes in the relic 
density can occur if one allows for the existence of phases in the 
gaugino mass parameters as shown elsewhere \cite{Nath,sasa} which can 
only happen if one departs from the minimal scenario. 

The method of partial waves presented here will be generalized to 
cover the subdominant processes of annihilation of the neutralino
 LSP's to channels other than that of a fermion pair and to include also 
the coannihilation processes which are important in particular 
regions of the parameter space. The results of such an analysis 
will be presented in a forthcoming publication.


\vspace*{0.6cm}
\noindent 
{\bf Acknowledgements} \\ 
\noindent 
M. Argyrou acknowledges support from IRAKLEITOS-Fellowship 
for Research of NKUA. 
A.B.L. acknowledges support from HPRN-CT-2000-00148
and HPRN-CT-2000-00149 programmes. He also thanks the University of Athens
Research Committee for partially supporting this work. 
D.V.N. acknowledges support by D.O.E. grant DE-FG03-95-ER-40917.
The work of  V.C.S. was supported in part
by DOE grant DE--FG02--94ER--40823.


\clearpage
\begin{thebibliography}{99}

\bibitem{EHNOS}
J.~Ellis, J.~S.~Hagelin, D.~V.~Nanopoulos, K.~A.~Olive
and M.~Srednicki, Nucl. Phys.  B238 (1984) 453; 
H.~Goldberg, Phys. Rev. Lett. 50 (1983) 1419.

\bibitem{wmap}
C.~L.~Bennett {\it et al.},
Astrophys.\ J.\ Suppl.\  {\bf 148} (2003) 1
[arXiv:astro-ph/0302207].


\bibitem{LNM}
For a review see, A.~B.~Lahanas, N.~E.~Mavromatos and D.~V.~Nanopoulos, 
Int. J. Mod. Phys. D 12 (2003) 1529.


\bibitem{eoss}
J.~R.~Ellis, K.~A.~Olive, Y.~Santoso and V.~C.~Spanos,
Phys.\ Lett.\ B {\bf 565} (2003) 176
[arXiv:hep-ph/0303043].


\bibitem{LN}
A.~B.~Lahanas and D.~V.~Nanopoulos,
Phys.\ Lett.\ B {\bf 568} (2003) 55
[arXiv:hep-ph/0303130].


\bibitem{cmssmmap}
H.~Baer and C.~Balazs,
JCAP {\bf 0305} (2003) 006
[arXiv:hep-ph/0303114];
U.~Chattopadhyay, A.~Corsetti and P.~Nath,
Phys.\ Rev.\ D {\bf 68} (2003) 035005
[arXiv:hep-ph/0303201];
C.~Munoz,
arXiv:hep-ph/0309346;
R.~Arnowitt, B.~Dutta and B.~Hu,
arXiv:hep-ph/0310103.



\bibitem{coanni1}
J.~R.~Ellis, T.~Falk and K.~A.~Olive,
Phys.\ Lett.\ B {\bf 444} (1998) 367; 
J.~R.~Ellis, T.~Falk, K.~A.~Olive and M.~Srednicki,
Astropart.\ Phys.\  {\bf 13} (2000) 181
[Erratum-ibid.\  {\bf 15}, 413 (2001)]
[arXiv:hep-ph/9905481].


\bibitem{coanni2}
M.~E.~Gomez,G.~Lazarides and C.~Pallis, 
Phys.\ Rev.\ D {\bf 61} (2000) 123512;
R.~Arnowitt, B.~Dutta and Y.~Santoso,
Nucl.\ Phys.\ B {\bf 606} (2001) 59
[arXiv:hep-ph/0102181];
T.~Nihei, L.~Roszkowski and R.~Ruiz de Austri, 
JHEP {\bf 0207} (2002) 024
[arXiv:hep-ph/0206266];
J.~Edsjo, M.~Schelke, P.~Ullio and P.~Gondolo,
JCAP {\bf 0304}, 001 (2003)
[arXiv:hep-ph/0301106].



\bibitem{hbrsb}
K.~L.~Chan, U.~Chattopadhyay and P.~Nath,
Phys.\ Rev.\ D {\bf 58} (1998) 096004
[arXiv:hep-ph/9710473].


\bibitem{focus}
J.~L.~Feng, K.~T.~Matchev and T.~Moroi,
Phys.\ Rev.\ Lett.\  {\bf 84} (2000) 2322
[arXiv:hep-ph/9908309];
J.~L.~Feng, K.~T.~Matchev and T.~Moroi,
Phys.\ Rev.\ D {\bf 61} (2000) 075005
[arXiv:hep-ph/9909334];
J.~L.~Feng, K.~T.~Matchev and F.~Wilczek,
Phys.\ Lett.\ B {\bf 482} (2000) 388
[arXiv:hep-ph/0004043].




\bibitem{Drees} M.~Drees and M.~Nojiri, Phys. Rev. {\bf D47} (1993) 376;
R.~Arnowitt and P.~Nath, Phys. Lett. {\bf B299} (1993) 58. 



\bibitem{LNS1} 
A.~B.~Lahanas, D.~V.~Nanopoulos and V.~C.~Spanos,
Phys.\ Rev.\ D {\bf 62} (2000) 023515;
Mod.\ Phys.\ Lett.\ A {\bf 16} (2001) 1229.


\bibitem{gannis}
J.~R.~Ellis, T.~Falk, G.~Ganis, K.~A.~Olive and M.~Srednicki,
Phys.\ Lett.\ B {\bf 510} (2001) 236
[arXiv:hep-ph/0102098].

\bibitem{baer}
H. Baer, M. Brhlik, M. Diaz, J. Ferrandis, P. Mercadante,
P. Quintana and X. Tata, Phys.\ Rev.\ D  {\bf 63} (2001) 015007.


\bibitem{LNS4} A.~B.~Lahanas, D.~V.~Nanopoulos and V.~C.~Spanos,
Phys.\ Lett.\ B {\bf 518} (2001) 94
[arXiv:hep-ph/0107151]; 
A.~B.~Lahanas, D.~V.~Nanopoulos and V.~C.~Spanos,
arXiv:hep-ph/0112134 and hep-ph/0211286.   


\bibitem{LS}
A.~B.~Lahanas and V.~C.~Spanos,
Eur. Phys. J. C23 (2002) 185.  


\bibitem{Kim}
M.~Drees, Y.~G.~Kim, T.~Kobayashi and M.~M.~Nojiri, 
Phys. Rev. D63 (2001) 115009.



\bibitem{x0} 
B.~Lee and S.~Weinberg,
Phys.\ Rev.\ Lett. {\bf 39}  (1977) 165.


\bibitem{x1} 
M.~Srednicki, R.~Watkins and K.A.~Olive,
Nucl.\ Phys.\ B {\bf 310}  (1988) 693.


\bibitem{x2}
J.~Ellis, D.V.~Nanopoulos, L.~Roszkowski 
and D.~N.~Schramm, Phys.\ Lett.\ B {\bf 245}  (1990) 251; 
J.~Ellis, L.~Roszkowski and Z.~Lalak, Phys.\ Lett.\ B {\bf 245}  (1990) 545; 
L.~Roszkowski Phys.\ Lett. \ B {\bf 252 } (1990) 471; 
Phys. Lett. B {\bf 262}(1991) 59;
J.~Ellis and L.~Roszkowski, Phys. \ Lett.\ B {\bf 283}  (1992) 252;
L.~Roszkowski and  R.~Roberts, Phys. \ Lett. \ B {\bf 309} (1993) 329;
G.L.~Kane, C.~Kolda, L.~Roszkowski and J.D.~Wells, 
Phys. \ Rev. \ D {\bf 49 } (1994) 6173.



\bibitem{x3} 
K.A.~Olive and M.~Srednicki, Phys. \ Lett.  \  B {\bf 230}(1989) 78;
Nucl. \ Phys. \ B {\bf 355} (1991) 208;
K.~Griest, M.~Kamionkowski and M.~S.~Turner, Phys. \ Rev. \ D {\bf 41}(1990) 3565;
J.~McDonald, K.~A.~Olive and M.~Srednicki, Phys. \ Lett. \ B {\bf 283} (1992) 80; 
S.~Mizuta, D.~Ng and M.~Yamaguchi, Phys. \ Lett. \ B {\bf 300 } (1993) 96.


\bibitem{x4}  J.~Ellis and F.~Zwirner, Nucl. \ Phys.  \ B {\bf 338 }(1990) 317;
M.~M.~Nojiri, Phys. \ Lett. \  B {\bf 261} (1991) 76;
M.~Kawasaki and S.~Mizuta, Phys. \ Rev. \ D {\bf 46} (1992) 1634.

\bibitem{x5}
J.~L.~Lopez, D.~V.~Nanopoulos and K.~Yuan,  Phys.\ Lett. \ B {\bf267 } (1991) 219;  
J.~L.~Lopez, D.~V.~Nanopoulos, 
H.~Pois and K.~Yuan, Phys. \ Lett. \ B {\bf 273}(1991) 423;
J.~L.~Lopez, D.~V.~Nanopoulos and K.~Yuan, 
Nucl. \ Phys. \ B {\bf 370} (1992) 445; Phys. \ Rev. \ D {\bf 48}(1993) 2766;  
S.~Kelley, J.~L.~Lopez, D.~V.~Nanopoulos, H.~Pois 
and K.~Yuan, Phys. \ Rev. \ D {\bf 47 } (1993) 2461.


\bibitem{x6} 
R.~Arnowitt and P.~Nath, Phys.\ Rev. \ Lett. \ {\bf 70} (1993) 3696;
Phys. \ Rev.  \ D {\bf 54} (1996) 2374;
M.~Drees and A.~Yamada, Phys. \ Rev. \ D {\bf 53} (1996) 1586;
J.~Ellis, T.~Falk, K.~A.~Olive and M.~Schmitt, Phys. \ Lett. \ B {\bf 388} (1996) 97.


\bibitem{x7} 
J.~Ellis, T.~Falk, G.~Ganis and K.A.~Olive,
Phys. \ Rev. \ D {\bf 58} (1998) 095002.

\bibitem{x8}
G.~Jungman, M.~Kamionkowski and K.~Griest,
Phys.\ Rept.\ {\bf 267} (1996) 195.


\bibitem{x9} 
K.~Griest, Phys.\ Rev.\ D {\bf 38} (1988) 2357 [Erratum: {\bf 39} (1989) 3802].
J.~Ellis and R.~Flores, 
Nucl.\ Phys.\ B {\bf 307} (1988) 833.
R.~Barbieri, M.~Frigeni and G.~Guidice,
Nucl.\ Phys.\ B {\bf 313} (1989) 725.
K.~Griest and D.~Seckel, Phys. \ Rev. \ D {\bf 43} (1991) 3191.
P.~Gondolo and G.~Gelmini, Nucl. \ Phys. \ B {\bf 360} (1991) 145.
A.~Bottino et. al., 
Phys.\ Lett.\ B {\bf 295} (1992) 330; 
M.~Drees and M.~M.~Nojiri, 
Phys.\ Rev.\ D {\bf 48} (1993) 3483; 
V.~A.~Bednyakov, H.~V.~Klapdor-Kleingrothaus and S.~Kovalenko,
Phys.\ Rev.\ D {\bf 50} (1994) 7128; 
P.~Nath and R.~Arnowitt,
Phys.\ Rev.\ Lett.\ D {\bf 74} (1995) 4592; 
E.~Diehl, G.~L.~Kane, C.~Kolda and J.~D.~Wells, 
Phys.\ Rev.\ D {\bf 52} (1995) 4223; 
L.~Bergstrom and P.~Gondolo, 
Astropart.\ Phys.\  {\bf 6} (1996) 263;
H.~Baer and M.~Brhlik, Phys. \ Rev. \ D {\bf 53 } (1996) 597;
M.~Drees, M.~M.~Nojiri, D.~Roy and Y.~Yamada,
Phys.\ Rev.\ D {\bf 56} (1997) 276; 
J.~Edsjo and P.~Gondolo,
Phys.\ Rev.\ D {\bf 56}, 1879 (1997)
[arXiv:hep-ph/9704361];
V.~Barger and C.~Kao, Phys. \ Rev. \ D {\bf 57} (1998) 3131.
H.~Baer and M.~Brhlik,
Phys.\ Rev.\ D {\bf 57} (1998) 567; 
J.~D.~Vergados, 
Phys.\ Rev.\ D {\bf 83} (1998) 3597; 
J.~Wells, Phys.\ Lett.\ B4 {\bf 43} (1998) 196.
M.~Brhlik, D.~J.~Chung and G.~L.~Kane,
Int.\ J.\ Mod.\ Phys.\ D {\bf 10} (2001) 367;
V.~A.~Bednyakov and H.~V.~Klapdor-Kleingrothaus, 
Phys.\ Rev.\ D {\bf 63} (2001) 095005; 
M.~E.~Gomez and J.~D.~Vergados, 
Phys.\ Lett.\ B {\bf 512} (2001) 252;
V.~D.~Barger and C.~Kao, 
Phys.\ Lett.\ B {\bf 518} (2001) 117;
L.~Roszkowski, R.~Ruiz de Austri and T.~Nihei, 
JHEP 0108 (2001) 024;
H.~Baer, C.~Balazs, A.~Belyaev, J.~K.~Mizukoshi, X.~Tata and 
Y.~Wang, JHEP 0207 (2002) 050;
U.~Chattopadhyay and P.~Nath,
Phys.\ Rev.\ D {\bf 66} (2002) 093001.

\bibitem{Belanger:2001fz}
G.~Belanger, F.~Boudjema, A.~Pukhov and A.~Semenov,
Comput.\ Phys.\ Commun.\  {\bf 149}, 103 (2002)
[arXiv:hep-ph/0112278]; 
G.~Belanger, F.~Boudjema, A.~Pukhov and A.~Semenov,
arXiv:hep-ph/0210327.



\bibitem{Baer:2002fv}
H.~Baer, C.~Balazs and A.~Belyaev,
JHEP {\bf 0203}, 042 (2002)
[arXiv:hep-ph/0202076].


\bibitem{Nihei}
T.~Nihei, L.~Roszkowski and R.~Ruiz de Austri,
JHEP {\bf 0105}, 063 (2001) [arXiv:hep-ph/0102308]; 
JHEP {\bf 0203} (2002) 031
[arXiv:hep-ph/0202009].

\bibitem{falk2}
T.~Falk, K.~A.~Olive and M.~Srednicki,
Phys.\ Lett.\ B {\bf 354} (1995) 99
[arXiv:hep-ph/9502401].


\bibitem{Haber}
H.~E.~Haber, hep-ph/9405376.

\bibitem{Kolb}
E.~W.~Kolb and M.~S.~Turner, ``The Early Universe", 
(Addison -- Wesley, N.Y. 1990).


\bibitem{mtop}
CDF Collaboration, D0 Collaboration and Tevatron Electroweak Working Group
arXiv:hep-ex/0404010.



\bibitem{edm}
E. Commins, et. al., Phys. Rev. {\bf A50}, 2960(1994); 
P.G. Harris et.al., Phys. Rev. Lett. {\bf 82}, 904(1999); 
S.~K.~Lamoreaux, J.~P.~Jacobs, B.~R.~Heckel, F.~J.~Raab and E.~N.~Fortson,
Phys.\ Rev.\ Lett.\  {\bf 57}, 3125 (1986).

\bibitem{deuteron}
D.~A.~Demir, O.~Lebedev, K.~A.~Olive, M.~Pospelov and A.~Ritz,
Nucl.\ Phys.\ B {\bf 680} (2004) 339
[arXiv:hep-ph/0311314]; 
O.~Lebedev, K.~A.~Olive, M.~Pospelov and A.~Ritz,
arXiv:hep-ph/0402023.


\bibitem{pilaf2}
A. Pilaftsis, Phys. Rev. {\bf D58}, 096010; Phys. Lett.{\bf B435}, 
88(1998);
~A. Pilaftsis and C.E.M. Wagner, Nucl. Phys. {\bf B553}, 3(1999);
~D.A. Demir, Phys. Rev. {\bf D60}, 055006(1999);
~S.~Y.~Choi, M.~Drees and J.~S.~Lee,
Phys.\ Lett.\ B {\bf 481}, 57 (2000);
~M.~Boz,
Mod.\ Phys.\ Lett.\ A {\bf 17}, 215 (2002).

\bibitem{ibrah}
T. Ibrahim and P. Nath,  
Phys.Rev.D63:035009,2001; hep-ph/0008237;
T.~Ibrahim, Phys.\ Rev.\ D {\bf 64}, 035009 (2001);
T.~Ibrahim and P.~Nath,
Phys.\ Rev.\ D {\bf 66}, 015005 (2002);
~S.~W.~Ham, S.~K.~Oh, E.~J.~Yoo, C.~M.~Kim and D.~Son,
arXiv:hep-ph/0205244.


\bibitem{Nath}
M.~E.~Gomez, T.~Ibrahim, P.~Nath and S.~Skadhauge, hep-ph/0404025.


\bibitem{sasa}
T.~Nihei and M.~Sasagawa, hep-ph/0404100.




\end{thebibliography}
\end{document}